\documentclass[10pt,amsmath,amssymb,aps,pre, twocolumn,letterpaper,floatfix,superscriptaddress]{revtex4-1}

\usepackage{amsmath}
\usepackage{graphicx, xcolor} 
 \usepackage{bm}
 \usepackage{epstopdf}
 \usepackage{textcomp}
 \usepackage{gensymb}
 \usepackage{makecell}
 \usepackage{tablefootnote}
 \usepackage{blindtext}
\usepackage[section]{placeins}

\graphicspath{{./figures/}}
\renewcommand{\(}{\left(}
\renewcommand{\)}{\right)}

\begin{document}\sloppy

\title{Diffusive search and trajectories on spatial networks: a propagator approach}

	 \author{Zubenelgenubi C. Scott}
	 \affiliation{Department of Physics, University of California, San Diego, San Diego, California 92093}
	
	 \author{Aidan I. Brown}
	 \affiliation{Department of Physics, Ryerson University, Toronto, Canada}
	
	 \author{Saurabh S. Mogre}
	 \affiliation{Department of Physics, University of California, San Diego, San Diego, California 92093}
	
	 \author{Laura M. Westrate}
	 \affiliation{Department of Chemistry and Biochemistry, Calvin University, Grand Rapids, Michigan 49546}

	 \author{Elena F. Koslover}
	 \email{ekoslover@ucsd.edu}
	 \affiliation{Department of Physics, University of California, San Diego, San Diego, California 92093}

\begin{abstract}
			Several organelles in eukaryotic cells, including mitochondria and the endoplasmic reticulum, form interconnected tubule networks extending throughout the cell.			
			These tubular networks host many biochemical pathways that rely on proteins diffusively searching through the network to encounter binding partners or localized target regions.
			Predicting the behavior of such pathways requires a quantitative understanding of how confinement to a reticulated structure modulates reaction kinetics.
			In this work we develop both exact analytical methods to compute mean first passage times and efficient kinetic Monte Carlo algorithms to simulate trajectories of particles diffusing in a tubular network. Our approach leverages exact propagator functions for the distribution of transition times between network nodes and allows large simulation time steps determined by the network structure. The methodology is applied to both synthetic planar networks and organelle network structures, demonstrating key general features such as the heterogeneity of search times in different network regions and the functional advantage of broadly distributing target sites throughout the network. The proposed algorithms pave the way for future exploration of the interrelationship between 
			tubular network structure and biomolecular reaction kinetics.
\end{abstract}
		
\maketitle

\section{Introduction}

Diffusive transport in geometrically complex environments underlies a broad variety of biophysical phenomena, ranging from transcriptional regulation in the nucleus~\cite{benichou2011facilitated,koslover2011theoretical},
to reactions inside organelle structures~\cite{lizana2009controlling,brown2020impact}, to intercellular communication through a variety of channel and bridge arrangements~\cite{zani2010cellular}.
 The morphology of the confining environment is known to fundamentally alter the kinetics of diffusion-limited chemical reactions, switching between compact and non-compact search processes depending on the effective dimensionality of the domain~\cite{benichou2010geometry, benichou2014first}. Multimolecular reaction systems such as phosphorylation cascades may acquire novel dynamic behaviors such as ultrasensitivity and bistability depending on the degree of confinement~\cite{takahashi2010spatio,abel2012membrane}.

A particularly important class of confined diffusion processes occurs on network structures, which have been used to describe porous media~\cite{Bouchaud1990anomalous,bryant1993physically}, neuronal trees~\cite{sartori2020statistical}, and organelle morphologies~\cite{viana2020mitochondrial,lin2014structure}. These `spatial networks' are characterized by nodes and edges embedded in physical space, a restriction which limits the network topology. Encompassing a broad variety of transport and communications networks, spatial networks have limited node degree, with each node connected only to a handful of neighbors in close physical proximity~\cite{barthelemy2011spatial}. There is an extensive body of literature on characterizing the behavior of random walks on general networks (see~\cite{masuda2017random} for a review). Many studies focus on systems where particles exhibit a well-defined hopping time across each edge, with hops treated either as discrete time steps~\cite{noh2004random, hwang2012first} or as constant-rate Poisson processes~\cite{dora2020active,wales2002discrete}. Others allow for generalized distributions of hop times that are nevertheless uniform throughout the network~\cite{montroll1965random,fouxon2016solvable}.
Recently, a general theory for heterogeneous continuous-time random walks on networks has been developed~\cite{grebenkov2018heterogeneous}, which incorporates transition times with arbitrary distributions that are specific to each node. A similar approach has previously been applied, modeling transitions on a network of states embedded in an energy landscape~\cite{koslover2012force}.

Particles diffusing along the edges of a network exhibit a broad range of inter-node transition times, whose distribution is dependent on the edge length. Prior work focusing on simple networks of tubes and containers~\cite{lizana2005diffusive}, as well as planar networks resembling percolation clusters~\cite{brown2020impact}, highlighted the importance of edge length in overall search times on the network. Here, we develop mathematical methodology for computing reaction mean first passage times (MFPT) on arbitrary spatial networks, with diffusion incorporated as a concrete physical transport process along each edge. Whereas past work on diffusive processes simplified individual node transitions to a single effective rate constant~\cite{dora2020active,wales2002discrete}, we explicitly incorporate the edge-length-dependent distribution of transition times, bridging local diffusive particle dynamics and large-scale transport on spatial networks.

For many reaction-diffusion systems, the behavior of interest requires quantities beyond the mean first passage time and other low-order moments of the first-passage distribution. For example, biochemical processes may rely on extreme value statistics that dictate the time-scales for the first of many signaling particles to reach a target~\cite{Schuss2019redundancy}. Activation cascades can be modulated by processive rebinding processes wherein one enzyme can return repeatedly to the same target after its first encounter~\cite{takahashi2010spatio}. Targets that themselves undergo diffusive motion are prevalent in most biomoleculear reaction systems. Modeling of these more complex processes requires moving beyond analytically tractable methods to leverage stochastic simulations of diffusing particles on network structures. 

A family of agent-based simulation methods, known as kinetic Monte Carlo~\cite{oppelstrup2009first} or Green's function reaction dynamics~\cite{vanzon2005green,van2005simulating} are particularly well suited for simulating sparse populations of particles diffusing in complex geometries. These methods rely on analytically determined propagator functions to evolve individual particles within `protective domains' -- regions where they do not come in contact with other particles. By sampling from an appropriate propagator, such simulations allow time steps that are tuned to the local structure of the domain, minimizing the computational time involved in propagating a particle through empty space. Several general propagators have been developed for three-dimensional regions~\cite{sokolowski2019egfrd}, and this approach has been employed for simulating systems such as kinase cascades~\cite{takahashi2010spatio}, target search by DNA-binding proteins~\cite{koslover2011theoretical}, and multi-modal transport of peroxisome vesicles in fungal hyphae~\cite{mogre2018multimodal}. In this work we develop exact analytical propagators for the passage of diffusing particles between network nodes. By allowing for step sizes comparable to the local edge length, these propagators enable simulations that can be run much more efficiently than classical Brownian dynamics. We thus present an efficient propagator-accelerated algorithm optimized for stochastic simulations of diffusing particles confined in spatial networks.

We focus specifically on tubular networks similar to those formed by intracellular reticulated organelles. The peripheral endoplasmic reticulum is one such network; it forms a dynamic web of interconnected tubules with a topologically continuous lumen, spread throughout the cell periphery\cite{westrate2015form}. The ER hosts a variety of biochemical reaction pathways and plays a crucial role in calcium dynamics, lipid delivery, and protein synthesis and quality control~\cite{Schwarz2016}. Another reticulated organelle structure is formed by the fusion of mitochondria in yeast and many mammalian cell types~\cite{rafelski2013mitochondrial,sukhorukov2012emergence}. Mitochondrial network structures share many topological features with geographical transportation networks~\cite{viana2020mitochondrial} and are thought to reside in the percolation regime, exhibiting just barely enough connectivity to form a large cell-spanning connected cluster~\cite{sukhorukov2012emergence}. The functional role of mitochondrial network formation remains a topic of much debate~\cite{hoitzing2015function}, but is thought to include complementation of mtDNA defects~\cite{busch2014quality}, quality control through selective fusion and mitophagy~\cite{twig2011interplay,patel2013optimal}, and enhanced energy transmission~\cite{skulachev2001mitochondrial}. 

In both the ER and mitochondrial networks, critical biomolecular reactions require individual diffusive components to find each other within the extensive network architecture. Some functions, such as the packaging of newly synthesized secretory proteins into ER exit sites or the regulation of transciption in mitochondrial nucleoids, rely on proteins reaching relatively stationary punctate target sites in the network. Although some recent evidence indicates the possibility of directed transport for ER luminal proteins driven by local fluid flow~\cite{holcman2018single}, the major form of transport within organelle networks is still believed to be diffusive.

In this work, we develop new algorithms for quantifying target search and reaction kinetics inside tubular network structures similar to those exhibited by reticulated cellular organelles. Specifically, our approach is well suited to spatial networks with well-defined edge lengths and low node degrees (typically 3 or less for ER~\cite{lin2014structure} and mitochondrial networks~\cite{viana2020mitochondrial}). Furthermore, we rely on the simplifying assumption that motion along tubular edges, rather than trapping at voluminous nodes, dominates diffusive transport times.
Our overarching goal is to be able to accurately compute the distribution of search and encounter times for diffusive particles within network morphologies. To this end, we employ both analytical methods to extract low-order moments (mean and variance) of the search times, as well as describing an efficient algorithm for agent-based stochastic simulations that can incorporate mutually reactive diffusing particles. The result is a mathematical framework optimally suited for modeling the kinetics of a broad variety of molecular processes confined in tubular networks.

\section{Model development and transition time distributions}
\label{sec:model}

We consider the diffusive motion of particles on a network embedded in physical space. Specifically, the network structure consists of point-like nodes ($i, j, \ldots$), connected by one-dimensional edges of length $\ell_{ij}$. The edges can be curved, and thus longer than the Euclidian distance between the connected nodes, but can only connect to other edges at a node.

\begin{figure}
	\includegraphics[width=0.4\textwidth]{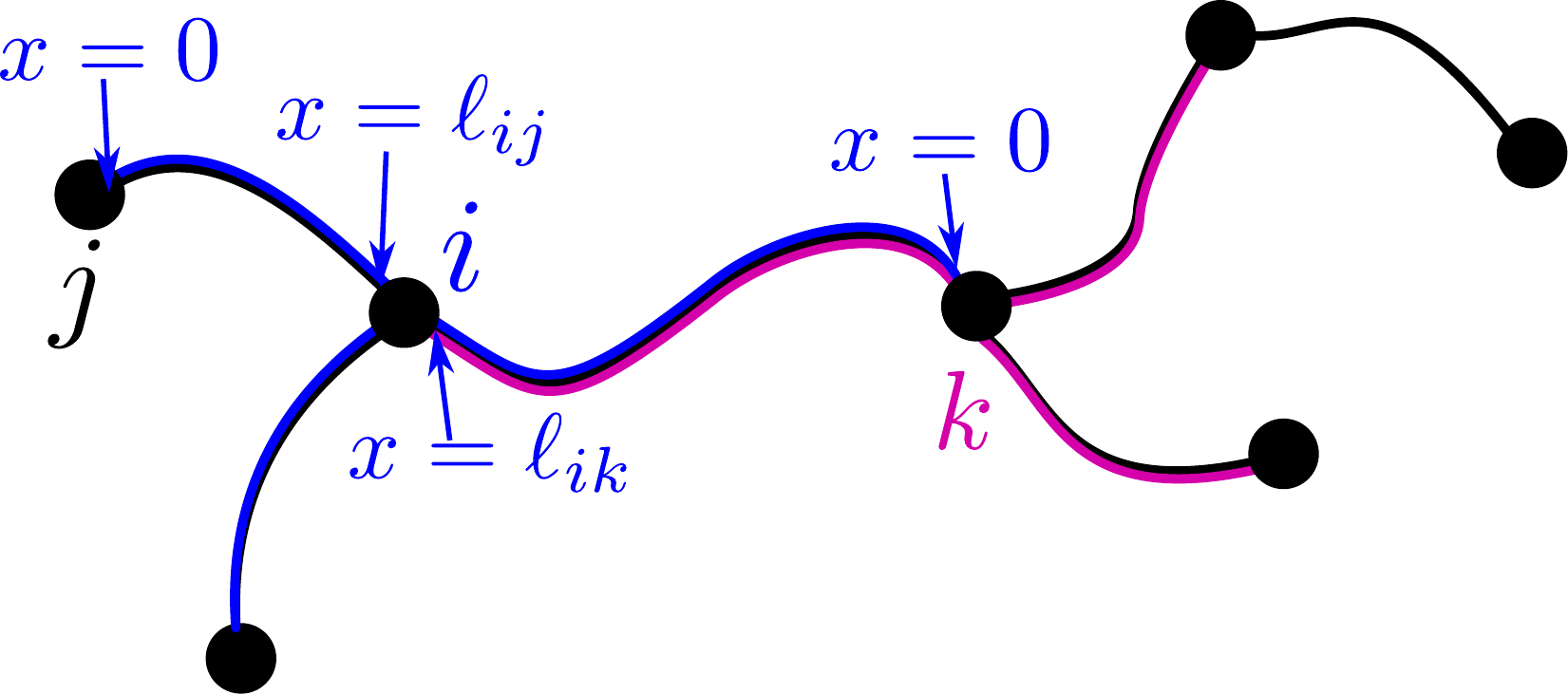}%
	\caption{Model schematic, showing small section of an example network. Blue lines show the neighborhood of node $i$; magenta lines show the neighborhood of node $k$. The coordinate system for two edges within neighborhood $i$ is labeled, with $x$ going from $0$ (at adjacent node) to $\ell_{ij}$ (at node $i$ itself).}%
	\label{fig:coordinatesystem}%
\end{figure}

Particles diffuse in one dimension along these edges, with diffusivity $D$. They do not spend any finite time trapped at the nodes themselves, which serve merely as point-like edge intersections. A particle that starts at such an intersection will diffuse around the edges adjacent to that node (a `node neighborhood') until it hits an adjacent node. It will then continue to diffuse in the neighborhood of the new node.
 We can thus consider the particle as moving from neighborhood to neighborhood with a certain distribution of transition times between adjacent neighborhoods. It is important to note that the current neighborhood of a particle is defined not by the edge on which it is located but rather by the last node that it has crossed (Fig.~\ref{fig:coordinatesystem}). 
With this definition, the motion of the particle can be treated as a Markov state model~\cite{chodera2014markov}, with each Markov state corresponding to a neighborhood. Transitions between neighborhoods are memoryless -- once a particle hits a given node, its subsequent distribution is no longer dependent on the previous nodes it has passed through. However, unlike prior applications of Markov state models~\cite{chodera2014markov} or `discrete path sampling'~\cite{wales2002discrete,lorenzo2020thermal} in the context of molecular rearrangements, the transition times between states are not Poisson distributed. Instead, this model falls in the category of heterogeneous continuous time random walks, which can be analyzed on arbitrary networks for arbitrary distributions of transition times~\cite{grebenkov2018heterogeneous}. 

To quantify the behavior of such a system, we need to know the local transition time distribution $P_{ik}(t)$, which gives the probability density that a particle which starts at node $i$ at time $0$ will first hit an adjacent node $k$ between time $t$ and $t+dt$, without first reaching any other nodes in the meantime. We note that in contrast to previous analyses of random walks on networks~\cite{masuda2017random,grebenkov2018heterogeneous}, the distribution of waiting times to leave the neighborhood and the splitting probability of which node is next encountered are not independent random variables ({\em i.e.}\ the conditional distribution of first passage times differs depending on which of the adjacent nodes is reached first). 
 To calculate the local transition time distributions, we generalize a well-known approach for  finding the flux of diffusive particles to the edges of a linear segment with absorbing boundaries~\cite{redner2001guide}.
 A one-dimensional coordinate system ($0\leq x \leq \ell_{ik}$) is placed along each edge attached to node $i$, with $x=0$ corresponding to node $k$ and $x=\ell_{ik}$ corresponding to the junction node $i$ where the particle starts (see Fig.~\ref{fig:coordinatesystem}).

Along each edge $ik$ of the neighborhood, the Green's function distribution $c_{ik}(x,t)$ of a diffusing particle obeys the usual diffusion equation:
\begin{subequations}
\begin{align}
\frac{\partial c_{ik}}{\partial t} & = D\frac{\partial^2 c_{ik}}{\partial x^2},\label{eq:diff1}\\
c_{ik}(x,0) & = \frac{1}{d_i} \delta\(x-\ell_{ik}\),\label{eq:diffinit}\\
c_{ik}(0,t) & = 0,\label{eq:diffbound}
\end{align}
\label{eq:diff}%
\end{subequations}
where $d_i$ is the degree of the $i^\text{th}$ node in the network. Equation~\ref{eq:diffinit} gives the initial condition, indicating that the particle distribution is concentrated in a Dirac-delta function at the junction node. Equation~\ref{eq:diffbound} gives the boundary conditions, with each neighboring node treated as an absorbing boundary. The transition time distibution function is then given by the flux into each of these boundaries:
\begin{equation}
\begin{split}
P_{ik}(t) = D \left. \frac{\partial c_{ik}}{\partial x}\right|_{x=0}.
\end{split}
\end{equation}
Integrating each $P_{ik}$ over time gives the probability that a particle will hit adjacent node $k$ before any other (ie: the splitting probability from node $i$). An additional boundary condition at each edge is set by enforcing continuity of the particle distribution function at the junction node: $c_{ik}(\ell_{ik},t) = c_{ij}(\ell_{ij},t)$, for all adjacent nodes $k, j$. 

Equation~\ref{eq:diff} can be solved by way of a Laplace transform $t\rightarrow s$, which gives the following equation for the transformed Green's function $\widehat{c}_{ik}$
\begin{equation}
\begin{split}
s\widehat{c}_{ik} - c_{ik}(t=0) = D \frac{\partial^2 \widehat{c}_{ik}}{\partial x^2}.
\end{split}
\label{eq:chat}
\end{equation}
The homogeneous solution is 
\begin{equation}
\begin{split}
\widehat{c}_{ik}(x,s) = A \sinh(\alpha x) \prod_{j\neq k} \sinh\(\alpha \ell_{ij}\),
\end{split}
\label{eq:cik}
\end{equation}
where $\alpha = \sqrt{s/D}$, and the product is over all edges attached to node $i$, other than edge $k$. The prefactor $A$ is set by the initial condition, and can be found by integrating Eq.~\ref{eq:chat} over an infinitesimally small interval around the junction node. This gives
\begin{equation}
\begin{aligned}
- 1 & = - D \sum_{k=1}^{d_i} \left. \frac{\partial \widehat{c}_{ik}}{\partial x}\right|_{\ell_{ik}},  \\
A & = \frac{1}{\alpha D} \( \prod_{j=1}^{d_i} \sinh \alpha \ell_{ij}\)^{-1} \(\sum_{k=1}^{d_i} \coth \alpha \ell_{ik} \)^{-1},
\end{aligned}
\label{eq:A}
\end{equation}
where the sums and products are over all edges attached to node $i$. Finally, the flux to the absorbing boundaries is given by
\begin{equation}
\begin{split}
\widehat{P}_{ik} =  \(  \sinh \alpha \ell_{ik} \sum_{j=1}^{d_i} \coth \alpha \ell_{ij}\)^{-1}
\end{split}
\label{eq:flux}
\end{equation}

Equation~\ref{eq:flux} gives the Laplace-transformed distribution of times for a particle starting at junction node $i$ to first reach adjacent node $k$ without hitting any other neighboring node. We can also define the survival time distribution $Q_{i}(t)$---the probability that the particle has not reached any of the neighboring nodes by time $t$. Its Laplace-transformed form is:
\begin{equation}
\begin{split}
\widehat{Q}_{i} = \frac{1 - \sum_{k=1}^{d_i} \widehat{P}_{ik}}{s} 
\end{split}
\label{eq:Q}
\end{equation}

Expanding around $s=0$ gives the splitting probability to each neighboring node ($P^{*}_{ik}$) and the overall average waiting time ($Q^{*}_{i}$) before the particle hits one of the neighboring nodes. Specifically, 
\begin{subequations}	
\begin{align}
P_{ik}^* & = \lim_{s\rightarrow 0} \widehat{P}_{ik}(s) = \frac{1/\ell_{ik}}{\sum_{j=1}^{d_i} 1/\ell_{ij}} \label{eq:P*}\\
Q_{i}^* & = \lim_{s \rightarrow 0} \widehat{Q}_i(s) = \frac{1}{2D}. \frac{\sum_{j=1}^{d_i} \ell_{ij}}{\sum_{j=1}^{d_i} 1/\ell_{ij}} \label{eq:Q*}
\end{align}
\label{eq:PQ*}%
\end{subequations}
The next higher-order term for small $s$ can be used to calculate the variance in the transition time to one of the neighboring nodes (see Sect.~\ref{sec:difflimrxn}).

The above expressions enable propagation of particles that start specifically on a node to the neighboring nodes. In many applications it is useful instead to consider particles that start distributed along the edges of the network. Such a particle must first be propagated to one of the nodes bounding that edge, after which its behavior can again be described by the node-to-node propagator (Eq.~\ref{eq:flux}.) For a particle starting at position $x_0$ along edge $m$, of length $\ell_m$, the Laplace-transformed flux to either of the two boundaries has the well-known form~\cite{redner2001guide}
\begin{equation}
\begin{split}
j_- = \frac{\sinh \left[\alpha(\ell_m-x_0)\right]}{\sinh\left[\alpha \ell_m\right]}, \quad
j_+ = \frac{\sinh \left[\alpha x_0\right]}{\sinh\left[\alpha \ell_m\right]}.
\end{split}
\label{eq:jedge}
\end{equation}

If the particle starts uniformly distributed along edge $m$,  the corresponding flux to each of the bounding nodes ($\widehat{P}^{(E)}_{mj}$) and the survival probability on the edge ($\widehat{Q}^{(E)}_m$) are given by 
\begin{equation}
\begin{split}
\widehat{P}^{(E)}_{mj} & = \frac{1}{\alpha\ell_m} \tanh \left(\frac{\alpha\ell_m}{2} \right) \\
\widehat{Q}^{(E)}_m  & = \frac{1}{s}\( 1 - \frac{2}{\alpha\ell_m} \tanh \(\frac{\alpha\ell_m}{2}\)\).
\end{split}
\label{eq:fluxE}
\end{equation}
Expanding around $s=0$ gives the trivial splitting probability for a particle starting uniformly on the edge: ${P}^{(E*)}_{mj} = 1/2$, and the average waiting time to leave the edge:
\begin{equation}
\begin{split}
\widehat{Q}^{(E*)}_m  & = \ell_m^2/(12 D)
\end{split}
\label{eq:QEavg}
\end{equation}

The splitting probabilities and survival times enable the calculation of mean first passage times on the network (Sect.~\ref{sec:MFPT}).
In order to sample from the full distribution of transition times, an inverse Laplace transform must be applied to the propagators in Eqs.~\ref{eq:flux} and~\ref{eq:fluxE}, as discussed in Sect.~\ref{sec:simulations}.

\section{Computing mean and variance of first passage times}
\label{sec:MFPT}

\subsection{Diffusion-limited reactions}
\label{sec:difflimrxn}
We next present an analytic approach for calculating low-order moments of the reaction time distribution for particles that react instantaneously upon reaching a set of target nodes in the network.
 We note that the derivation in this section largely reiterates previously published work~\cite{grebenkov2018heterogeneous, koslover2012force,brown2020impact}, but is presented here for completeness and consistency of notation.

The probability that a particle starting at node $i$ at time $0$ is in the neighborhood of node $j$ at time $t$ is defined by $G_{ij}(t)$. 
 The Laplace-transformed form of this propagator has been previously derived, both for general continuous-time random walks on networks~\cite{grebenkov2018heterogeneous} and for specific applications involving diffusive motion of particles in interconnected tubules~\cite{brown2020impact} or over multi-state energy landscapes~\cite{koslover2012force,lorenzo2020thermal}. It can be expressed as
\begin{equation}
\begin{split}
\widehat{G}_{ij} = \left[\(\mathbf{I} - \mathbf{\widehat{P}} \)^{-1} \right]_{ij} \widehat{Q}_j,
\end{split}
\end{equation}
where the elements of matrix  $\mathbf{\widehat{P}}$ are given by Eq.~\ref{eq:flux} if two nodes are connected by an edge in the network, and are zero otherwise. The propagator $G_{ij}(t)$ gives the probability that the particle last hit node $j$ sometime before time $t$ and has not yet left the neighborhood of that node. 

To calculate the distribution of first passage times to any set target of nodes $\{k\}$ in the network, we treat those nodes as being perfect absorbers. That is, whenever the particle first hits node $k$, it instantaneously vanishes from the network. The case of finite reaction rates in localized network regions is treated separately in the next section.
We remove all rows and columns corresponding to the target neighborhoods from the matrix $\mathbf{\widehat{P}}$ as well as the vector $\widehat{\overrightarrow{Q}}$. 
As a result, the time a particle spends in the neighborhood of any network node is not altered, but when the particle leaves that neighborhood by moving to a target node, it is removed entirely from the network rather than continuing to propagate further~\cite{koslover2012force,brown2020impact}. The survival probability that a particle starting at node $i$ has not left the network by time $t$ is $H_i(t) = \sum_{k=1}^N G_{ik}(t)$, where the summation is over all nodes on the network. For particles initially distributed over nodes, with $V_i$ the probability of starting at node $i$, the survival probability is given by the following matrix expression:
\begin{equation}
\begin{split}
\widehat{H}(s) = \overrightarrow{V} \cdot \(\mathbf{I} - \mathbf{\widehat{P}} \)^{-1} \cdot \widehat{\overrightarrow{Q}}.
\end{split}
\label{eq:Hs}
\end{equation}
The central inverted matrix is a normalized form of a weighted discrete Laplacian over the network, which is used in a broad class of problems involving random walks on networked structures~\cite{grebenkov2018heterogeneous,masuda2017random}.

\begin{figure*}
	\includegraphics[width=\textwidth]{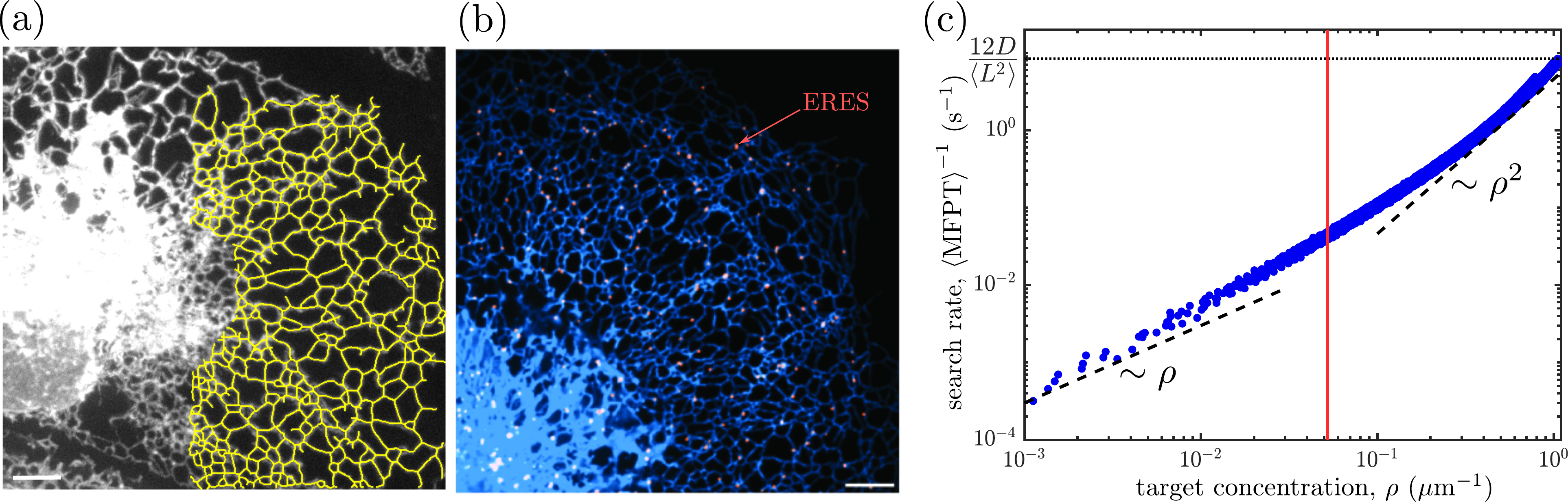}
	\caption{Target search rate on peripheral ER networks. (a) Network structure (yellow) extracted from confocal image of the COS7 cell peripheral ER network. (b) Peripheral ER network (blue) and ER exit sites (red, marked by fluorescently labeled Sec23a, a COPII protein). 
		Scale bars in (a) and (b) are $5\mu$m. (c) Dependence of search rate (inverse of mean first-passage time) on target concentration (per tube length) in nine distinct ER network structures. Each dot shows the average over 100 realizations of a fixed number of target nodes uniformly distributed over a single network structure. Black dashed lines indicate appropriate scaling for a two-dimensional ($\sim \rho$) and one-dimensional ($\sim \rho^2$) continuum. Horizontal dotted line shows limiting case of average search rate on a network composed entirely of target nodes. Red line gives ERES density extracted from (b). Particles are assumed to have a diffusivity of $D=1\mu\text{m}^2/s$.
	}
	\label{fig:targetdensity}
\end{figure*}

A natural extension is to consider particles starting on the edges of the network, with probability $W_m$ of starting (uniformly distributed) along edge $m$. In this case, the Laplace-transformed survival probability can be expressed as
\begin{equation}
\begin{split}
\widehat{H}^{(E)}(s) = \overrightarrow{W} \cdot \left[ \widehat{\overrightarrow{Q}}^{(E)} + \mathbf{P}^{(E)} \cdot\(\mathbf{I} - \mathbf{\widehat{P}} \)^{-1} \cdot \widehat{\overrightarrow{Q}} \right].
\end{split}
\label{eq:HsE}
\end{equation}
Here, the first term represents particles that never leave their initial edge and the second term includes a propagator for moving from the edge to one of its bounding nodes, convolved with the propagators for moving across all node-to-node paths through the network, and finally the survival probability of remaining at some final node neighborhood.
 Columns of $\widehat{\mathbf{P}}^{(E)}$ corresponding to target nodes are again removed from the matrix. The elements of $\widehat{\mathbf{P}}^{(E)}$ and $\widehat{\overrightarrow{Q}}^{(E)}$ are given in Eq.~\ref{eq:fluxE}.

Regardless of whether particles start on nodes or edges of the network, the mean first passage time  to encounter the set of targets is given by
\begin{equation}
\begin{split}
\tau = \lim_{s\rightarrow 0} \widehat{H}(s),
\end{split}
\end{equation}
which can be evaluated directly with the aid of Eq.~\ref{eq:PQ*}a,~\ref{eq:PQ*}b, and~\ref{eq:QEavg}. Similarly the variance in the time to find a target is given by
\begin{equation}
    \sigma^2 = \langle \tau^2\rangle - \langle \tau \rangle^2,
\end{equation}
where
\begin{equation}
    \langle \tau^2 \rangle = - 2 \left. \frac{\partial \widehat{H}}{\partial s}\right|_{s=0} .
\end{equation}
From this, the mean square first passage time is expressed as
\begin{equation}
\begin{aligned}
    \langle \tau^2 \rangle = -2 \overrightarrow{V}\cdot & \left[ \(\mathbf{I} - \mathbf{\widehat{P}} \)^{-1} \cdot \frac{\partial \mathbf{\widehat{P}}}{\partial s} \cdot \(\mathbf{I} - \mathbf{\widehat{P}}  \)^{-1} \cdot \widehat{\overrightarrow{Q}} \right. \\ %
    &\left. + \(\mathbf{I} - \mathbf{\widehat{P}} \)^{-1} \frac{\partial\widehat{\overrightarrow{Q}}}{\partial s}\right],
\end{aligned}
\end{equation}
where the derivatives of $\mathbf{\widehat{P}}$ and $\widehat{\overrightarrow{Q}}$ are
\begin{subequations}
\begin{align}
    \left.\frac{\partial \widehat{P}_{ik}(s)}{\partial s}\right|_{s=0} &= -\frac{1}{6D}\(\frac{ \ell_{ik}}{\sum_j 1/\ell_{ij}} +\frac{ 2/\ell_{ik} \(\sum_j \ell_{ij}\)}{\(\sum_j 1/\ell_{ij}\)^2}\) \\
    \left. \frac{\partial \widehat{Q}_{i}(s)}{\partial s} \right|_{s=0} &= -\frac{1}{24D^2}\(\frac{ \sum_j \ell_{ij}^3}{\sum_j 1/\ell_{ij}} + \frac{ 8\(\sum_j \ell_{ij}\)^2 }{\(\sum_j 1/\ell_{ij}\)^2}\).
\end{align}
\end{subequations}

\subsubsection{Example: target search times in the endoplasmic reticulum}
As an example application of the calculations above, we consider network structures extracted from confocal images of the peripheral endoplasmic reticulum in COS7 cells. A data set of 9 peripheral ER images, obtained as described in prior work~\cite{brown2020impact},
  was used to extract tubular network structures (Fig.~\ref{fig:targetdensity}a; details in~\ref{app:networks}). 
   For these biologically important tubular networks, we consider how the distribution of times to find target nodes varies with the target concentration. This question is particularly important in the context of the early secretory pathway. Proteins destined for secretion are co-translationally inserted into the ER membrane or lumen, undergo folding and quality control~\cite{ellgaard2003quality,brown2021design}, and must leave the ER through punctate ER exit sites (ERES). These ERES are largely immobile sites scattered throughout the network~\cite{stadler2018diffusion} (Fig.~\ref{fig:targetdensity}b) and proteins are assumed to diffuse to one of these sites for capture and packaging into vesicles that enable them to leave the ER and proceed to further steps of secretory processing~\cite{budnik2009er,borgese2016getting}. 

It is interesting to consider what ERES density is sufficient to enable diffusing proteins to rapidly encounter exit sites.
 In a two- or three-dimensional continuum, 
reaction rates are proportional to the concentration of the target. However, in a geometry that is less than two-dimensional, the usual assumption of mass-action kinetics ceases to hold, and we expect a steeper dependence of rates upon target concentration~\cite{benichou2010geometry}.

For each individual ER network structure, we randomly distribute different numbers of target nodes across the network and compute the mean first passage time (MFPT) for a diffusive particle to first hit a target. The particles are assumed to start uniformly distributed over the edges of the network, with $W_m = \ell_m / \sum_n \ell_n$ the probability of starting on edge $m$.
In Fig.~\ref{fig:targetdensity}c we plot the search rate, defined as the inverse of the mean first passage times averaged over many choices of target node location.
We see the search rate is linearly proportional to the target concentration $\rho$ when the concentration is low, so that the search process occurs on an effectively two-dimensional structure. 
By contrast, when target density is high, the search rate becomes proportional to $\rho^2$, as would be expected in a one-dimensional system. This limit is reached when $1/\rho$ becomes comparable to the typical edge length of the ER network structures: $\ell \approx 1.2 \pm 0.1 \mu\text{m}$  (expected value of starting edge length for uniformly distributed particles, with standard error of the mean computed over different networks). In this limit, particles need only diffuse along a single edge before encountering a target site. 

It should be noted that there is a broad transition region between the two scaling regimes~\cite{mogre2020physical}, and that physiological estimates of exit site density
 fall within this intermediate region (red line in Fig.~\ref{fig:targetdensity}c). Thus, the rate at which proteins find ER exit sites in the peripheral network should be super-linearly dependent on the ERES concentration, so that increasing the number of sites can substantially speed up the encounter process.

\subsubsection{Variability of arrival times}
\begin{figure}
	\includegraphics[width=0.5\textwidth]{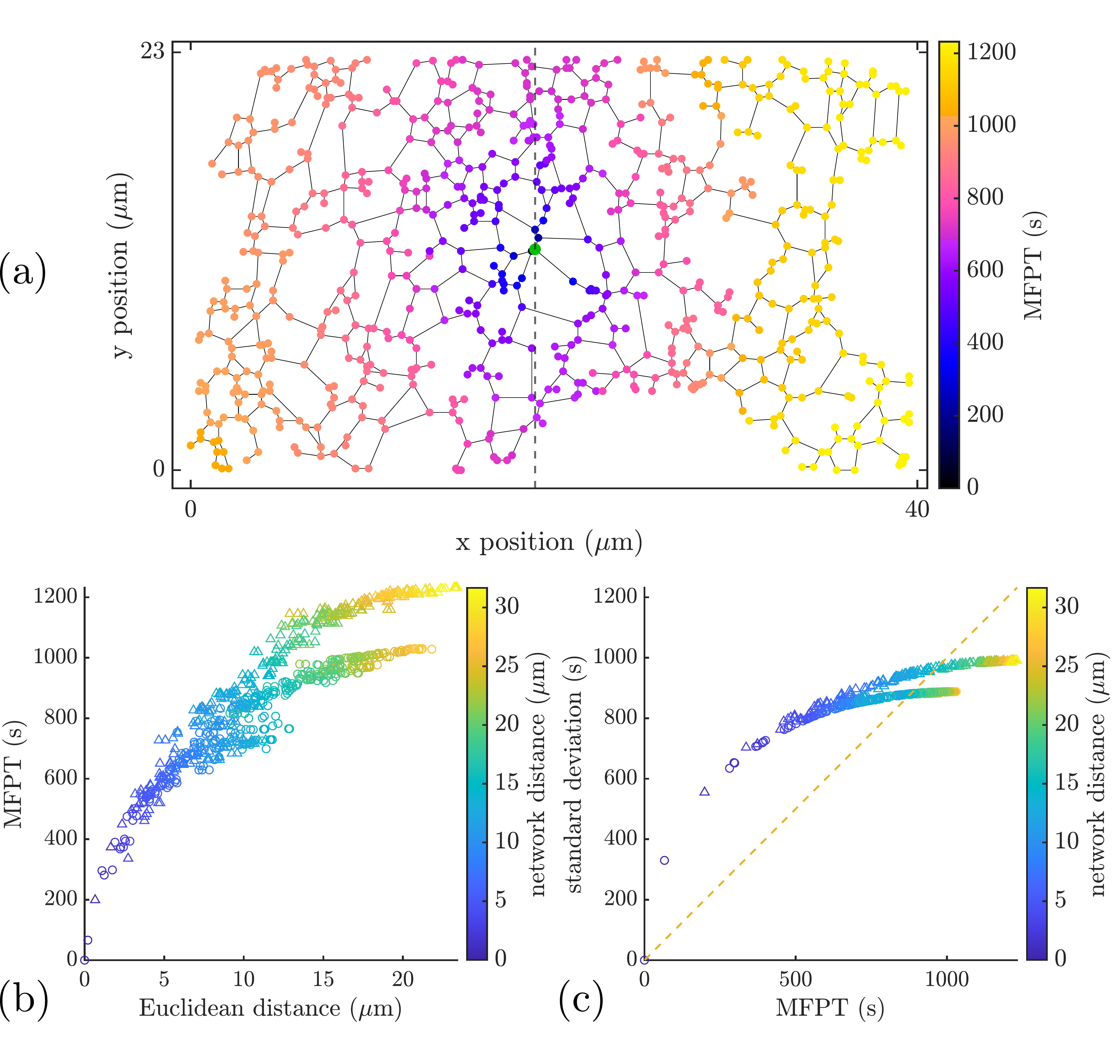}
	\caption{Plot of mean first-passage time (MFPT) versus starting distance from target. (a) ER network structure with a single target (green) and color indicating the MFPT to the target for particles starting at each node. (b) Dependence of MFPT on Euclidean distance from the target for each node; color indicates network distance of each node to the target; circles and triangles indicate nodes to the left and right of the target, respectively. 
		(c) Analogous plot showing standard deviation for the first passage time distribution. Dashed line shows standard deviation for a constant-rate Poisson process ($\sigma_\text{FPT} = \text{MFPT}$) for comparison. A diffusivity of $D=1\mu\text{m}^2/s$ is assumed.}%
	\label{fig:variance}%
\end{figure}

One important question in considering kinetics on complex geometries is the extent to which the mean first passage time can be used to characterize the full distribution of reaction times. For compact diffusive search on fractal geometries of dimension less than two, the first passage time distribution is known to exhibit a broad range of relevant time-scales, so that the search process is not well-characterized by the MFPT~\cite{grebenkov2018strong,benichou2010geometry}. Although the search for very sparse targets in ER networks appears to be effectively two-dimensional, signatures of geometry-controlled kinetics (such as a strong dependence on source and target position~\cite{benichou2010geometry}) are nevertheless observed. In particular, the mean first passage time varies substantially depending on the distance of the starting node from the target (Fig.~\ref{fig:variance}). Nodes at similar spatial (Euclidean) distances can also exhibit very different mean first passage times, due to the heterogeneous nature of the ER network connectivity. Because diffusing particles explore many paths from the source to the target, the MFPT can also be very different for nodes with similar values of the `network distance', defined as the shortest distance between two points measured along the network edges.
Furthermore, the standard deviation of the arrival time from a given starting node 
can be substantially larger than the mean first passage time itself, particularly for nodes located close to the target (Fig.~\ref{fig:variance}c). This effect again arises from the multiple time-scales associated with particles following a variety of different paths to the target site.

These observations imply that diffusion-limited reactions within the ER network deviate from the expectations of bulk kinetics, where arrival times generally represent a Poisson process with a well-characterized MFPT and a comparable standard deviation. Instead, particles starting in regions nearby and well-connected to the target arrive much faster than particles from far away. Furthermore, even for a given starting point, some particles travel rapidly directly to the target, while others meander away to explore the rest of the network, leading to broadly distributed first passage times.

\subsection{Finite reaction rates}

First passage times to perfectly absorbing network nodes represent purely diffusion-limited kinetics, where a reaction occurs as soon as the particle finds its target. A biologically relevant complication to this model would include finite reaction rates in certain regions of the network. Such rates become relevant when a reaction requires particles to undergo rearrangements, rotations, or conformational changes in addition to simply finding each other diffusively. If those rearrangements occur on a timescale that is relatively slow compared to the time for the reactants to diffusively separate again, then the reaction kinetics cannot be simply treated as first passage to a perfectly absorbing target site. Instead, the target site is assumed to have a particular rate of reacting with the particle, that is applicable only when the particle is within some minimal contact distance from the target. Models with spatially localized finite reaction rates have previously been employed in quantifying the kinetics of multi-conformation DNA-binding proteins searching for their genomic target sites~\cite{kochugaeva2017optimal}, and of vesicles encountering cytoskeletal filaments to initiate motor-driven transport~\cite{ando2015cytoskeletal}.
For one-dimensional models of a tubular geometry, finite reaction rates can also be used as a simplification to account for the radial diffusion time required to find a target by a particle that approaches its axial position~\cite{mogre2020hitching}.

In our model of  diffusion on tubular networks, particles spend all their time on network edges, with nodes serving only as intersections that allow transitions between edges. We therefore consider reaction rates that are associated with each edge of the network, defining $\gamma_{ij}$ as the reaction rate on the edge connecting nodes $i$ and $j$. Reactions near a particular node can be represented by setting the reaction rates in all edges around that node. If necessary, additional degree-2 nodes can be inserted along the edge to confine the reactive area still further. 

To solve for the mean reaction time in this model, we first consider the propagation of the particle from a single neighborhood ($i$). The Laplace-transformed probability distribution on each edge around node $i$,  $\widehat{c}_{ik}$, obeys a modified form of Eq.~\ref{eq:chat}. Namely,
\begin{equation}
\begin{split}
s\widehat{c}_{ik} - c_{ik}(t=0) = D \frac{\partial^2 \widehat{c}_{ik}}{\partial x^2} - \gamma_{ik} \widehat{c}_{ik}.
\end{split}
\label{eq:chat_finrate}
\end{equation}
This can be solved to find the flux into each of the adjacent nodes:
\begin{equation}
\begin{split}
\widehat{P}_{ik} =  \alpha_{ik} \( \sinh \alpha_{ik} \ell_{ik} \sum_{j=1}^{d_i} \alpha_{ij} \coth \alpha_{ij} \ell_{ij}\)^{-1},
\end{split}
\label{eq:flux_finrate}
\end{equation}
where $\alpha_{ij} = \sqrt{(s+\gamma_{ij})/D}$. The splitting probability $P_{ik}^*$ of hitting an adjacent node before any reaction occurs is given by plugging $s=0$ into the expression above. The probability of reacting before leaving the neighborhood of node $i$ can then be calculated as $1 - \sum_k P_{ik}^*$. 

The Laplace-transformed probability that the particle is still in the neighborhood by a certain time (having neither reacted nor reached an adjacent node) is given by
\begin{equation}
\begin{split}
\widehat{Q}_{i} = \sum_{k=1}^{d_i} \int_0^{\ell_{ik}} \widehat{c}_{ik} dx = \frac{1}{D} \frac{\sum_{j=1}^{d_i} \frac{1}{\alpha_{ij}} \tanh(\frac{\alpha_{ij}\ell_{ij}}{2})}{\sum_{j=1}^{d_i} \alpha_{ij} \coth(\alpha_{ij} \ell_{ij})}.
\end{split}
\label{eq:Q_finrate}
\end{equation}
Evaluating this expression at $s=0$ gives the average waiting time to leave the neighborhood.

Similarly, for particles starting uniformly distributed along network edge $m$, we can find the flux into one of the bordering nodes $j$ (defined as $\widehat{P}^{(E)}_{mj}$) and the survival probability on the edge ($\widehat{Q}^{(E)}_{m}$) according to
\begin{equation}
\begin{split}
\widehat{P}_{mj}^{(E)} &= \frac{1}{\alpha_m\ell_m} \tanh\(\frac{\alpha_m\ell_m}{2}\) \\
\widehat{Q}^{(E)}_m &= \frac{1}{\alpha_m^2 D}\left[ 1 - \frac{2}{\alpha_m \ell_m} \tanh\(\frac{\alpha_m\ell_m}{2} \) \right].
\end{split}
\end{equation}

These expressions can then be plugged directly into Eq.~\ref{eq:Hs} or Eq.~\ref{eq:HsE} to compute the mean first passage time to leave the network through either the finite-rate reactions along network edges or through reaching a perfectly absorbing target node. To find the MFPT in the presence of perfectly absorbing targets, all elements corresponding to the target nodes should be removed from the matrix expressions as before, so that reaching the targets is treated as permanently leaving the network.

\begin{figure*}
	\includegraphics[width=0.9\textwidth]{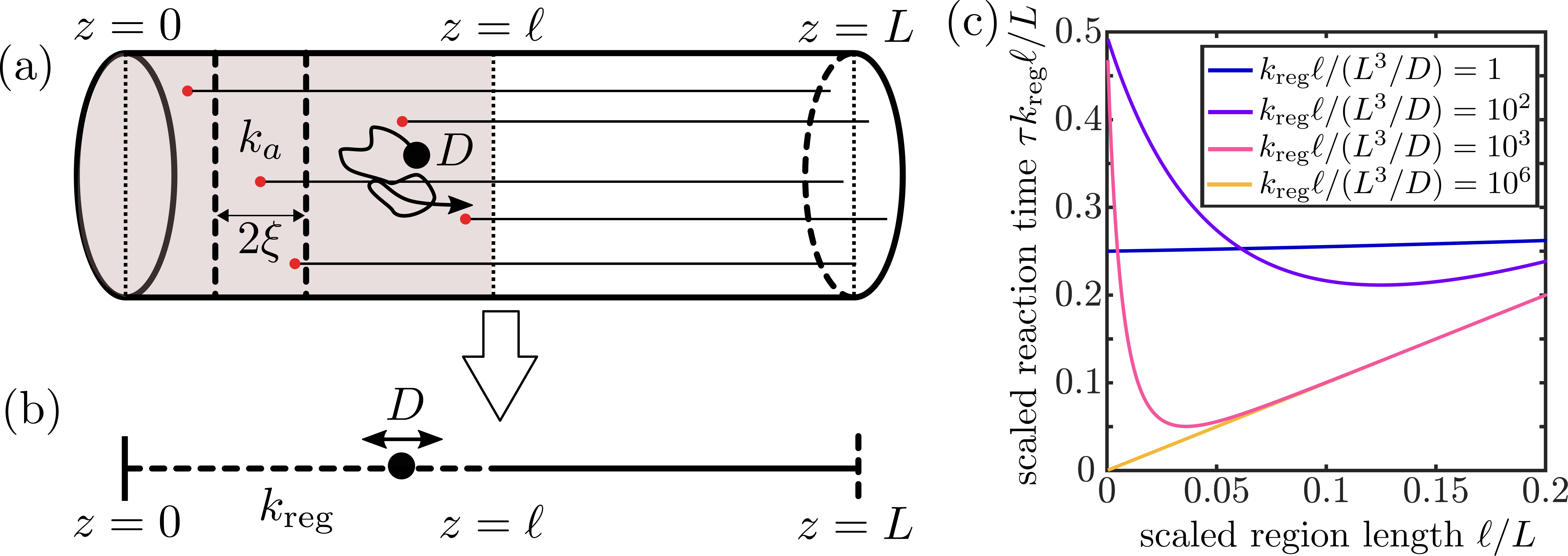}
	\caption{Reaction times for localized finite reactive region in one dimension. (a) Schematic of cylindrical domain representing a cellular projection (such as a neuronal axon), with reactive points representing microtubule tips distributed over a region of length $\ell$, with $k_a$ the absorbance rate for a particle that reaches an axial position within distance $\xi$ of a tip. (b) The cylindrical system is mapped to a one-dimensional model, with effective reaction rate $k_\text{reg}$ in the region where microtubule tips are distributed, a reflecting boundary at $z=0$ representing the distal tip of the projection, and an absorbing boundary at $z=L$ representing the cell body.		
		 (c) Scaled mean absorption time for particles originating at $z=0$ and diffusing in the partially reactive one dimensional domain. }	
	\label{fig:1Dabsorb}
\end{figure*}

\subsubsection{Example: Extended absorbing region in one dimension}

A simple example of the calculations described above involves particles on a one-dimensional interval containing a region with a finite reaction rate $k_\text{reg}$. 
Such a system can be thought of as a simplified representation of a tubular cellular domain, such as a fungal hypha or neuronal axon~\cite{mogre2020physical}. Organelles such as signaling endosomes and autophagosomes that are produced at the distal end of this domain must be loaded onto microtubules to be delivered in a retrograde fashion to the nuclear region~\cite{steinberg2014endocytosis,maday2014axonal}. There is evidence that some cellular cargos begin their retrograde journeys by binding preferentially to microtubule plus-end tips, which accumulate high concentrations of dynein motors and associated activator proteins to form a `loading zone' at the distal cell tip~\cite{lenz2006dynein,moughamian2013ordered}. An interesting question then arises regarding how the distribution of microtubule tips near the distal end affects the overall rate of loading the organelles. If all tips are localized right at the distal end, particles originating at the distal end will have a chance to bind as soon as they are formed. However, any particles that diffuse axially past the tips may end up exploring the full domain over long time periods without returning to the distally localized tips. On the other hand, a broader distribution of tip positions may make particles more likely to latch on before diffusing away. 

We explore this trade-off by mapping the distribution of microtubule tips in a cylindrical domain to an effective one-dimensional model (Fig.~\ref{fig:1Dabsorb}a, b), where $\ell$ represents the length of the region over which  microtubule tips are distributed. The overall attachment rate in this region can be approximated as $k_\text{reg} =  n k_a \xi/\ell$, where $n$ is the number of microtubule tips, $\xi$ is the contact radius for binding a tip, and $k_a$ is an effective binding rate that incorporates rapid radial diffusion to encounter the tip while at the appropriate axial position. We assume a reflecting boundary at $z=0$ representing the distal end, and an absorbing boundary at $z=L$ representing the region near the nucleus that serves as the target for the particles. Particles are initiated at $z=0$ and diffuse with diffusivity $D$ until they are either absorbed in the reactive region or reach the soma through diffusion alone. For simplicity, we neglect the motor-driven transport time to reach the soma after loading on the microtubules, focusing instead on the optimal dispersion of microtubule tips to minimize the loading time.
For a given number of microtubules, $k_\text{reg}\ell$ is expected to be constant and we explore how distribution over different region lengths $\ell$ affects the first passage time to loading. 

Treating the 1D simplified system (Fig.~\ref{fig:1Dabsorb}b) as a network with only two edges, one of which is absorbing, we plot (in Fig.~\ref{fig:1Dabsorb}c) the mean first passage time for a particle to either react with the microtubule tips or reach the far end of the domain. Interestingly, an optimum is observed with respect to the length $\ell$ over which the absorbing tips are distributed, indicating that it is advantageous to spread out microtubule tips in the distal region rather than placing them all as near as possible to the distal end. The optimum value of $\ell$ increases as the overall reactivity goes down.
When the tip reaction rate $k_a$ is very rapid, then $\ell_\text{opt} \rightarrow 0$ as no particle can make it past the most distal tips. By contrast, when $k_a$ is very low, the optimum disappears entirely as particles have the chance to explore the entire domain and reach the distal absorbing boundary without ever binding in the absorbing region, regardless of its size.

\begin{figure*}
	\includegraphics[width=\textwidth]{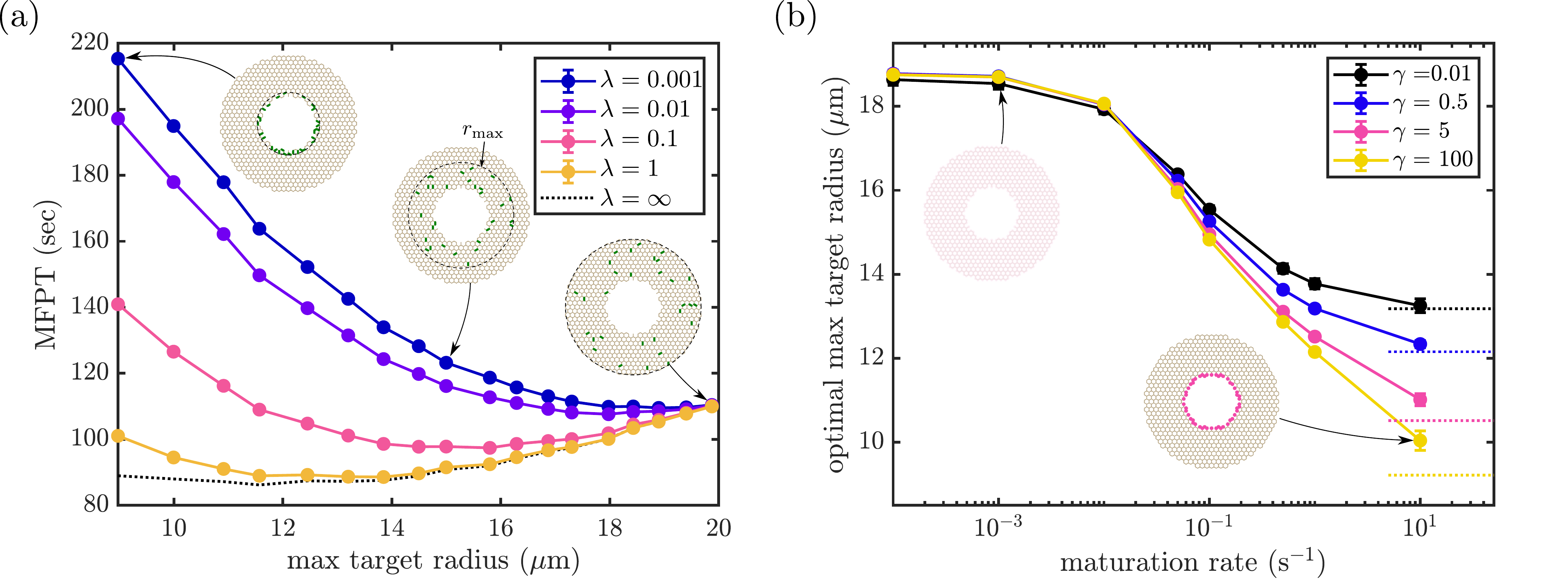}
	\caption{Mean first passage times on synthetic honeycomb networks with dimensions similar to the ER in COS7 cells. (a) The relationship between MFPT of particles initiated on interior nodes and maximum radial distance of 30 absorbing edges ($\gamma =1\text{s}^{-1}$) for various maturation rates, $\lambda$. Each dot shows the MFPT averaged over 500 different combinations of absorbing edges within the maximum target distance (errorbars included but not visible at this scale.) Example networks are shown for  small, medium and large maximum target distance, with target edges highlighted in green.
	Dashed black line shows the case with no maturation process ($\lambda = \infty$).
	(b) The effect of maturation rates, $\lambda$, on the optimal maximum target distance, with various absorption rates, $\gamma$. Dashed lines indicate optimal maximum target distance in limiting case of instantaneous maturation ($\lambda = \infty$).
	Inset networks illustrate particle distribution at time of maturation for two different choices of maturation rate.}
	\label{fig:donut}
\end{figure*}

\subsubsection{Example: Target distribution on a 2D network}%
\label{sec:donuts}

Similar to the one-dimensional case described in the previous section, we can explore how the distribution of reactive regions in a two-dimensional network such as the peripheral endoplasmic reticulum affects diffusive search times. Here, we use a synthetic network consisting of a honeycomb lattice structure in a circular band with dimensions comparable to the peripheral ER in COS7 cells (Fig.~\ref{fig:targetdensity}, insets). The network has an edge length $0.8\mu$m (equal to average edge length for ER networks used in this study), an inner radius of $8\mu$m representing the nucleus, and an outer radius of $20\mu$m representing the cell boundary.

Most protein synthesis in the ER is thought to occur in the ribosome-studded perinuclear sheets~\cite{westrate2015form}, so we consider particles that are initiated at the innermost nodes of the network. Localized reactive regions are distributed over the network edges to represent ER exit site structures. A finite reaction rate on each exit-site edge accounts for any additional process that a particle must undergo after reaching an exit site before it can be stably captured. Such processes could include rotation, molecular rearrangement, or entry into a narrow-necked ERES structure~\cite{borgese2016getting}. 

One question of particular interest is whether there is any functional advantage to scattering ERES throughout the peripheral network, rather than concentrating them in the perinuclear region where proteins are initially translated. Given that both luminal and membrane proteins have been shown to penetrate throughout the peripheral tubular structure~\cite{westrate2015form}, one possible advantage to well-dispersed ERES is to efficiently capture proteins that happen to diffuse deeply into the periphery. We therefore consider how the average reaction time varies when a fixed number of reactive sites are distributed across regions of different width surrounding the interior boundary where particles are initiated. 

Interestingly, unlike the one-dimensional case in Fig.~\ref{fig:1Dabsorb}, there is very little advantage to dispersing reaction regions over a broader region of the two-dimensional lattice network. The dashed black line in Fig.~\ref{fig:donut}a shows the MFPT for a network with $30$ edges with reaction rate $\gamma=1\text{s}^{-1}$, spread out over increasingly broad regions of the network. While there is a slight optimum when the reactive edges are allowed to spread to a radial distance of $12\mu$m, the difference between this system and one where targets are placed in the innermost region of the network (in the same location where particles originate) is less than $3\%$. Higher values of reactivity $\gamma$ make it even more advantageous to concentrate all reactive sites closer to the inner radius, while lower values of $\gamma$ make the system independent of the reactive edge distribution (data not shown), much as in the one-dimensional case. In no case does there appear to be a substantial advantage to spreading out the reactive edges. 

This result underscores a fundamental distinction between a well-connected 2D network and a 1D interval. In the network, moving target sites out from the central region where particles are initiated necessitates spreading those sites out over a longer band, leaving gaps for particles to be able to diffuse through without hitting any target. Particularly in the case of high reactivity $\gamma$, such gaps allow escape of particles into the periphery that would not be possible if the targets were concentrated near the inner radius of the network. This effect counterbalances the advantage to capturing particles which do manage to disperse through the network. As a result, there is no substantial benefit to placing target sites further out in the periphery for the 2D network case.

However, the dispersed placement of reactive sites can greatly enhance kinetics in the case where newly synthesized particles are incapable of reacting immediately upon production, as discussed in the subsequent section.

\subsubsection{Example: Maturing particles}

Another potentially important complication for a variety of biomolecular search processes is the existence of a `maturation time,' during which particles diffuse through the domain but are not able to undergo the relevant reactions. For example, newly manufactured proteins in the ER must undergo folding and quality control processes that can take from minutes to hours before they are able to be loaded at ER exit sites for export~\cite{hebert2007inandout,brown2021design}.
Such particle maturation can be easily incorporated in our model as  a Poisson process of rate $\lambda$ that occurs simultaneously with diffusion. While the particle is maturing, it is not capable of reacting even if it reaches the reactive edges. As soon as the particle matures, reactivity is enabled and the particle's mean first passage time can be determined from its  maturation point using an extension of the usual techniques outlined above.

Specifically, we define $\psi_{m}(x,t)$ as the overall density of particles on edge $m$ at time t. Its Laplace-transformed form is given by
\begin{equation}
\begin{split}
    \widehat{\overrightarrow{\psi}}_{m}(x,s) = \overrightarrow{V}\cdot \(\mathbf{I}-\mathbf{\widehat{P}} \)^{-1} \cdot \mathbf{\widehat{c}}\(x,s\),
\end{split}
\end{equation}
where the entries $\widehat{c}_{im}\(x,s\)$ contain the Laplace transformed particle densities along edge $m$ defined for the node neighborhood $i$, as given by Eq.~\ref{eq:cik}.

This approach allows for nonuniform particle densities along each edge. In order to find mean first passage time after maturation, we need to compute the overall survival probability, $H^{(E)}_{m}(x,t)$, for a particle starting at position $x$ on edge $m$. Such a particle can remain on the edge until time $t$, react during that time, or hit either of the bounding nodes ($i,j$) before time $t$. In the latter case, its survival probability can be obtained by convolving with the expression in Eq.~\ref{eq:Hs} for particles initiated on a node. The overall Laplace-transformed survival probability is obtained as
\begin{equation}
\begin{split}
    \widehat{H}^{(E)}_{m}(x,s) = \ &\widehat{Q}_{m}^{(E)}(x,s+\gamma_{m}) + \widehat{P}^{(E)}_{m,i}(x,s+\gamma_{m})\widehat{H}_{i}(s) \\%
    &+ \widehat{P}^{(E)}_{m,j}(x,s+\gamma_{m})\widehat{H}_{j}(s).
\end{split}
\end{equation}
Here, $\widehat{Q}_{m}^{(E)}(x,s)$ and $\widehat{P}^{(E)}_{m,i}(x,s)$ are the survival probability and flux to the bounding node, respectively, given the particle starts at $x$. These quantities can be obtained using the standard solutions for diffusion on a one-dimensional interval with two absorbing boundaries~\cite{redner2001guide} ({\em i.e.:} from Eq.~\ref{eq:jedge}). 

The overall mean first passage time after maturation (not including the maturation time of $1/\lambda$ itself) is then
\begin{equation}
\begin{split}
    \tau(\lambda) = \lambda \sum_{m} \int_0^{\ell_{m}} \widehat\psi_{m}(x,\lambda) \widehat{H}^{(E)}_{m}(x,s=0) dx,
\end{split}
\end{equation}
where the sum is over all edges in the network.

Again considering the synthetic honeycomb networks, we can investigate the effect of particle maturation on mean first passage time, and probe the functional advantage of widely-distributed ER exit sites. In Fig.~\ref{fig:donut}a we see that by adding a maturation process (\emph{e.g.}\ $\lambda \leq 0.1 \text{s}^{-1}$), the MFPT of particles starting at the inner boundary exhibits a  pronounced minimum when the reactive regions are spread out over a broader region of the network. As the maturation rate becomes slower (\emph{e.g.}\ $\lambda=0.001 \text{s}^{-1}$,) the particles have time to spread uniformly across the network before maturing and it becomes advantageous to disperse targets over the entire network structure.

The interplay of maturation and local reactivity is highlighted in Fig.~\ref{fig:donut}b. For lower maturation rates, it is best to distribute reactive regions throughout the network, regardless of local reaction rate $\gamma$. For higher maturation rates, we see the optimal distance vary depending on reactivty. For rapid absorption ($\gamma=100 \text{s}^{-1}$) concentrating targets near the inner boundary is advantageous, as most particles can be captured shortly after maturation before they have a chance to explore the network. For slower reactivity, many particles have a chance to explore the full network structure and placing the target regions more broadly dispersed throughout the network becomes advantageous.

These results highlight how the placement of reactive regions in a tubular network can lead to non-trivial effects on the overall reaction rate. The analytic approach described here allows for rapid, easily implemented calculations of the mean and variance of reaction times. This approach thus makes practical an extensive exploration of how network morphology and the distribution of target positions and/or reaction rates modulate kinetic processes involving stationary network structures.

\begin{figure*}
	\includegraphics[width=\textwidth]{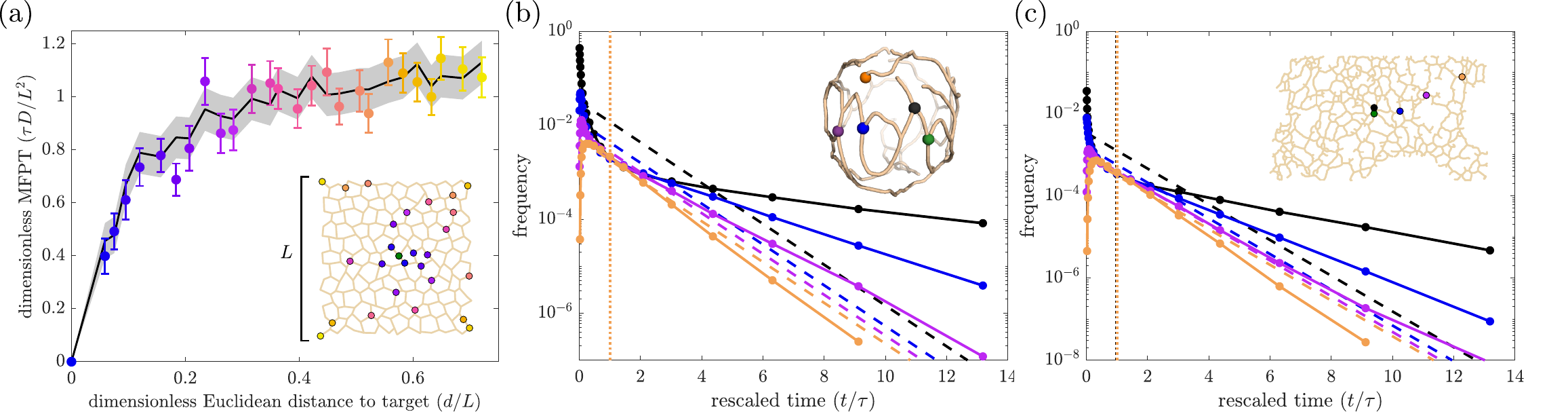}
	\caption{First passage times to a stationary target obtained with exact kinetic Monte Carlo simulations. (a) Mean first passage times for particles starting at different source nodes on a synthetic planar network to reach the target (green), plotted versus Euclidean distance from source to target. Error bars give standard error of the mean ($n=150$ particles simulated). Black line and shaded region give mean and expected standard error from corresponding analytical calculations. Diffusivity is set to $D=1\mu\text{m}^2/s$; length units are non-dimensionalized by network size $L$ and time units by $L^2/D$. (b) Distribution of first passage times to a target node (green), scaled by the MFPT ($\tau$), on a yeast mitochondrial network. Color corresponds to different starting nodes. Dashed lines show expected exponential decay for a Poisson process with the same mean. (c) Analogous plot for particles diffusing on a peripheral ER network structure.}
	\label{fig:simvalidate}
\end{figure*}

\section{Simulating particle trajectories}
\label{sec:simulations}
The analytic first passage time calculations described in the previous section are limited in several important aspects. First of all, they directly provide only the low-order moments (mean and variance) of first passage time distributions, without enabling the exploration of more detailed features, such as the extreme statistics for the earliest and latest arriving particles, which are important in a variety of biological signalling processes~\cite{Schuss2019redundancy}. In the complex, compact geometry of a reticulated network, the relevant arrival timescales can be very broadly distributed, and are not necessarily well-characterized by the mean and variance alone~\cite{grebenkov2018strong,benichou2010geometry}. Furthermore, our calculations thus far have been restricted to immobile targets or reactive regions within the network. Modeling the behavior of more complex reaction-diffusion processes, which may involve interaction between multiple mobile particles, can be achieved by switching from analytically tractable methods to agent-based stochastic simulations.

Agent-based simulations are particularly relevant when the reacting molecules are present in small copy numbers and substantial fluctuations of local concentration make mean-field descriptions of particle concentrations inaccurate. The simplest such reaction-diffusion simulations rely on Brownian dynamics, with particles taking very small discrete time steps and checking for a reaction on each step~\cite{andrews2004stochastic}. A series of alternate schemes have been developed to minimize the computational cost of propagating particles when they are far away from each other or the domain boundaries. Such methods, which have been variously termed  Green's Function Reaction Dynamics or first-passage kinetic Monte Carlo~\cite{vanzon2005green,van2005simulating,oppelstrup2009first,sokolowski2019egfrd}, leverage analytically computed propagator functions to move individual particles over large time steps within a `protected domain' where interaction with other particles or structures does not occur. They thus enable rapid propagation in regions where particle motion is purely diffusive, interspersed with a finer time-resolution in regions where reactions may take place. The main required assumption is that each propagated step of the particle is Markovian (memory-less) and that there are no reactions, obstacles, or interactions with other particles while the particle is propagating within the protected region.

Kinetic Monte Carlo approaches all rely on sampling the time interval for the next event of interest to occur, from an appropriate distribution of transition times. Original schemes relied on exponentially distributed transition times~\cite{gillespie1977exact}, while the more recent approaches sample from the analytically known distributions of first passage times in simple domains such as a one-dimensional interval or three-dimensional sphere~\cite{oppelstrup2009first,sokolowski2019egfrd}. Here, we leverage our diffusive propagator between network node neighborhoods (Sect.~\ref{sec:model}) to develop an efficient kinetic Monte Carlo simulation algorithm optimized for diffusive particles on tubular networks. In this section, we focus on instantaneous reactions with perfectly absorbing target nodes, although the method can easily be expanded to include finite reactivity on edges.

\subsection{Single particle propagator}

For a single particle propagating on a network, we consider the particle starting at node $i$ at time $t=0$. A protected domain can then correspond to the local neighborhood, containing all the edges connected to node $i$. The distribution of times to leave this domain is given by the flux into all the neighboring nodes, which are treated as absorbing boundaries. Specifically, we begin by sampling which neighboring node $k$ is first reached by the particle, from the discrete splitting probability  $P_{ik}^*$ given in Eq.~\ref{eq:P*}. The distribution of times to first hit this node, conditional upon not previously leaving the neighborhood, can then be obtained by the Laplace inversion of the flux $\widehat{P}_{ik}$ (Eq.~\ref{eq:flux}), normalized by $P_{ik}^*$. 

To perform the Laplace inversion, we first numerically calculate the poles of $\widehat{P}_{ik}$. These poles occur at discrete values $s_p$, which fall on the negative real axis and can be expressed as $s_p = - D u_p^2$, where the $u_p$ satisfy the following equation
\begin{equation}
\begin{split}
\sin\(\ell_{ik}u_p\)  \sum_{j=1}^{d_i} \cot \(\ell_{ij} u_p\)  = 0.
\end{split}
\label{eq:roots}
\end{equation}

The residues $r_p^{(ik)}$  at the poles can be found by evaluating the derivative of Eq.~\ref{eq:roots} with respect to $s$, yielding
\begin{equation}
\begin{split}
\frac{1}{r_p^{(ik)}} & = \frac{\sin \(\ell_{ik} u_p\)}{2Du_p} \sum_{j=1}^{d_i} \ell_{ij} \csc^2 \(\ell_{ij} u_p\).
\end{split}
\end{equation}

Finally, we evaluate the inverse Laplace tranform using the standard Bromwich integral together with the Cauchy residue theorem~\cite{arfken1999mathematical}:
\begin{equation}
\begin{split}
P_{ik}(t) = \sum_p r_p^{(ik)} e^{-Du_p^2 t}.
\end{split}
\label{eq:Pikinvert}
\end{equation}

In a kinetic Monte Carlo simulation, we initiate a particle on node $i$, sample the next neighboring node to be reached ($k$) according to probabilities $P_{ik}^*$, and then use inverse transform sampling to select the time interval $\Delta t$ to first reach this node according to the normalized probability distribution $\Delta t \sim P_{ik}(\Delta t)/P_{ik}^*$. Details of the sampling procedure are provided in~\ref{app:sampling}. To validate the simulations, we evaluate the mean and standard deviation in first passage times from different starting positions to a particular target on a 2D lattice-like network, showing a close match to analytical predictions (Fig.~\ref{fig:simvalidate}a).

\subsubsection{Examples: first passage time distributions}
Beyond the low-order moments, simulations allow us to explore the full distribution of first passage times to a target node. Figure~\ref{fig:simvalidate}b, c shows such distributions from several different starting nodes on an example yeast mitochondrial network~\cite{viana2020mitochondrial} and a peripheral ER network from a COS7 cell. These distributions are compared to the exponential distribution expected for a constant-rate Poisson process with the corresponding average time. For particles starting far from the target (orange curves), there is a peak ``most likely" time of target encounter. Shorter search times are precluded by the need to traverse a substantial distance in the network before ever hitting the target.  Particles that start near the target (black curves) exhibit long tails in the first passage time distribution, which are a known signature of compact search processes~\cite{benichou2010geometry}. These tails highlight the multiple timescales that contribute to the search, as some particles move towards the target immediately while others follow long meandering paths over the network before reaching the target. The deviation from a Poisson distribution, for both nearby and distant starting locations, indicate that reaction kinetics on these networks should have qualitatively different behaviors than would be expected in a bulk three-dimensional continuum.

\begin{figure}
	\includegraphics[width=0.5\textwidth]{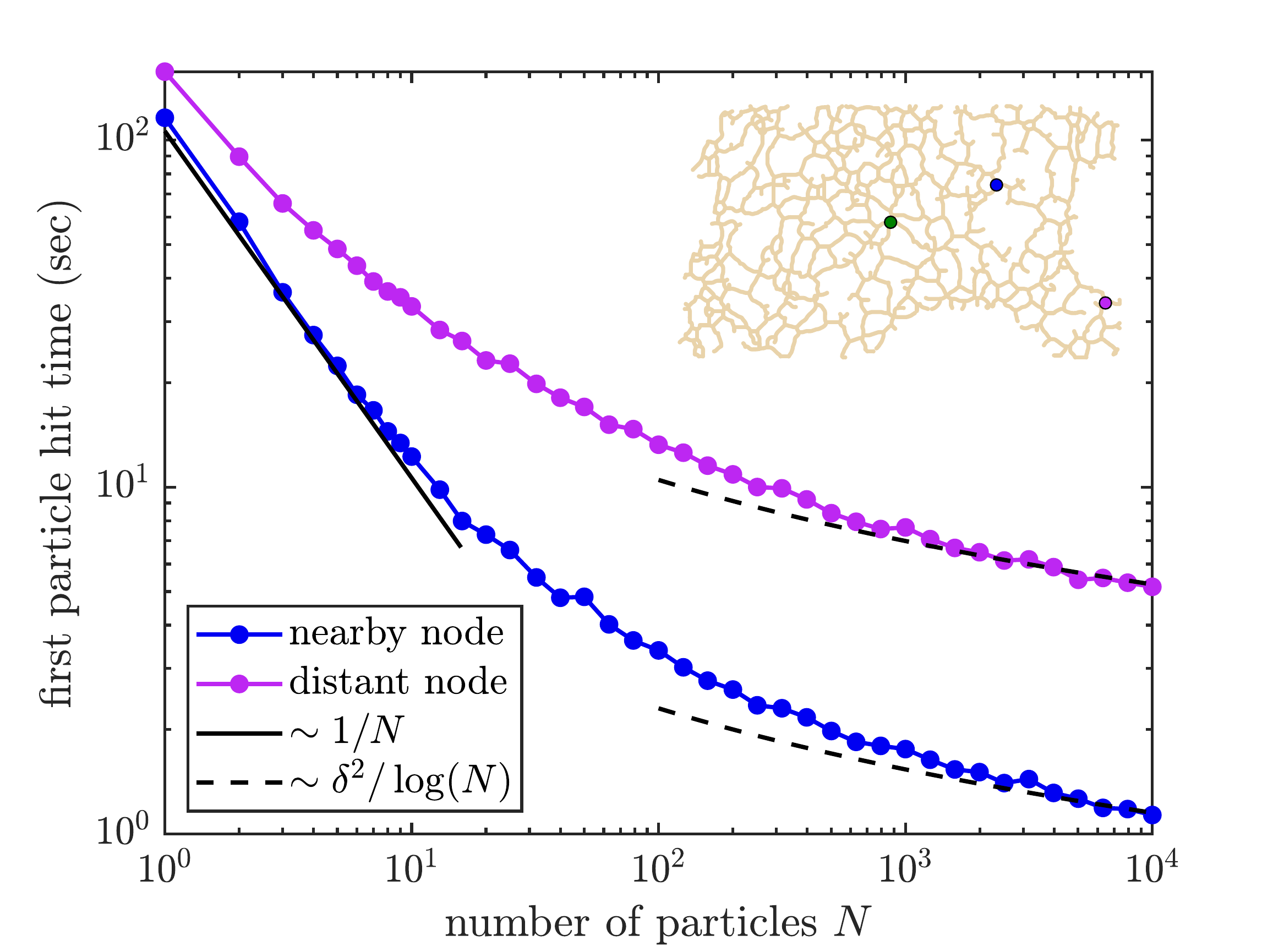}
	\caption{Extreme arrival statistics on an ER network structure. Plotted is the average first passage time for the first of $N$ particles to hit a target node (green), starting from either a nearby source (blue) or a distant source (magenta). Scaling laws are shown for the small $N$ limit (solid, $\sim 1/N$) and large $N$ limit (dashed, $\sim \delta^2/\log N$, where $\delta$ is the minimal network distance between start node and target). Diffusivity is set to $D=1\mu\text{m}^2/s$.}
	\label{fig:extremevals}
\end{figure}

Another application of stochastic particle hopping simulations is to explore the extreme arrival time---the average time for the first of many particles to find a target on the network. Such calculations are relevant to signaling processes where one or a few molecules are sufficient to trigger a response, so that the shortest rather than the mean arrival times set the timescale of signal initiation~\cite{Schuss2019redundancy}. 
 In Fig.~\ref{fig:extremevals}, we run simulations of $N$ independent particles starting from a given position on the network and show the time for the first of these particles to hit a particular target node. For a small number of simulated particles, the average first hitting time scales roughly as $\sim 1/N$, as would be expected in a bulk continuum. In other words, the reaction rate is approximately proportional to the particle concentration. However, for large numbers of particles, we observe a $\delta^2/\log N$ dependence, where $\delta$ is the minimal distance along the network between the starting point and the target. These results are
consistent with extreme first passage statistics in one-dimensional or two-dimensional systems~\cite{yuste1996order,basnayake2019asymptotic}, and are similar to previous simulations of particle hopping on networks which assumed a single timescale for each individual hop~\cite{dora2020active}.

\subsubsection{Synchronizing trajectories}

In certain applications, it is desirable to calculate snapshots of the particle position at prespecified times. For example, such trajectory information is needed to compute the mean squared displacement of a particle over different time intervals, a common analysis tool for quantifying the rate of diffusive spread. In this case, during each step of the kinetic Monte Carlo procedure, we check whether the time point $t+\Delta t$ exceeds the next save-time $t_s$. If that is the case, the particle needs to be propagated first over a time interval $\delta t = t_s - t$. To achieve this, we leverage a spatial `no-passage' propagator within the neighborhood of node $i$, which gives the distribution of particle positions conditional on not leaving the neighborhood by time $\delta t$. For a particle that starts on node $i$ at time $0$, and does not leave the neighborhood, the probability that it is on the edge connecting $i$ to $k$, between position $x$ and position $\ell_{ik}$, at time $\delta t$ is defined as $Y_{ik}(\delta t)$. The Laplace transform of this spatial cumulative distribution is:
\begin{equation}
\begin{split}
\widehat{Y}_{ik}(x,s) & = \int_x^{\ell_{ik}} \widehat{c}_{ik}(x',s) dx' \\ 
& = \frac{1}{s} \widehat{P}_{ik} \left[ \cosh(\alpha \ell_{ik}) - \cosh(\alpha x) \right]
\end{split}
\label{eq:Yik}
\end{equation}

To invert the Laplace transform, we note that this expression has the same poles as the flux $\widehat{P}_{ik}$, so that
\begin{equation}
\begin{split}
Y_{ik}(x,\delta t) & = \sum_p \frac{r_p^{(ik)}}{Du_p^2} \left[\cos(u_px) - \cos(u_p \ell_{ik}) \right]e^{-Du_p^2 \delta t}
\end{split}
\label{eq:Yikinv}
\end{equation}

To propagate the particle forward over a time interval $\delta t$, we first sample which edge $ik$ the particle is positioned on, using the discrete probabilities $Y_{ik}(0,\Delta t)$. The position along the edge is then sampled according to the conditional cumulative distribution: $\mathcal{P}(x'>x) =  Y_{ik}(x,\Delta t)/ Y_{ik}(0,\Delta t)$.

Once the particle has been placed at position $x_0$ along a specific edge $m$ (bounded by nodes $i$ and $k$), we can use a Laplace inversion of Eq.~\ref{eq:jedge} to sample the next time interval $\Delta t$ for the particle to first hit either node $i$ or node $k$. If this time interval would again exceed the next desired save point, the particle is propagated spatially along the edge  (see~\ref{app:sampling} for details).

\begin{figure}
	\includegraphics[width=0.5\textwidth]{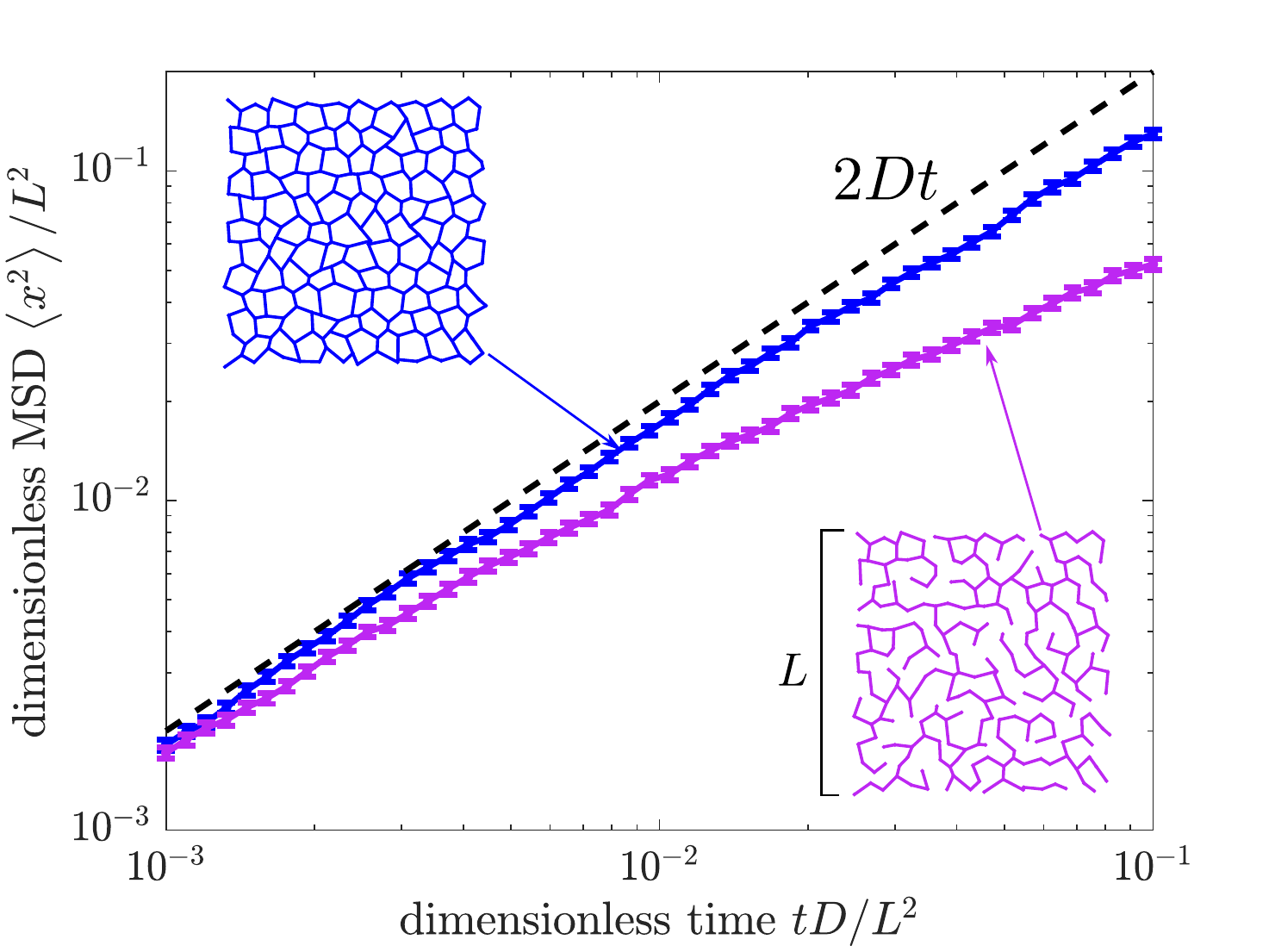}
	\caption{Mean squared displacement for simulated diffusing particles on two synthetic networks. Blue: Synthetic planar network with full lattice-like connectivity. Magenta: decimated network network with 27\% of edges removed while maintaining a single connected component. Length units are non-dimensionalized by network size $L$ and time units by $L^2/D$.}
	\label{fig:MSDsims}
\end{figure}

Simulations of complete particle trajectories can be used to explore the mean squared displacement (MSD) of particles diffusing over a network structure. As seen in Fig.~\ref{fig:MSDsims}, particles on a poorly connected network with many dead ends exhibit a lower, apparently subdiffusive, MSD than particles on a lattice-like network. This result implies that MSD analysis alone may indicate subdiffusive motion even when the particles are in fact undergoing classic diffusive motion, due to confinement on a network structure with missing connectivity.

\subsection{Simulating multiple particles}
Many problems concerning reaction kinetics on networked geometries require the ability to simulate not just individual particles searching for stationary target sites, but also  multiple mobile particles that encounter and react with each other. Additional complications arise in the multi-particle case because the possibility of reaction with a neighbor alters the distribution of each particle after any given time interval. This issue was addressed in previous studies by introducing `protective domains' that break up the available space into disjoint regions wherein each particle propagates independently of the others with no chance of interaction~\cite{oppelstrup2009first,van2005simulating}. 
For example, on a simple 1D interval, such protective domains can be defined by placing boundaries at all points half-way between the positions of neighboring particles. This notion was refined further by noting that when two particles approach close to each other, infinitesimal time steps can be avoided by defining a single protective domain surrounding both of them and then jointly propagating the pair distribution of the two particles~\cite{oppelstrup2009first}. 

\begin{figure}
	\includegraphics[width=0.5\textwidth]{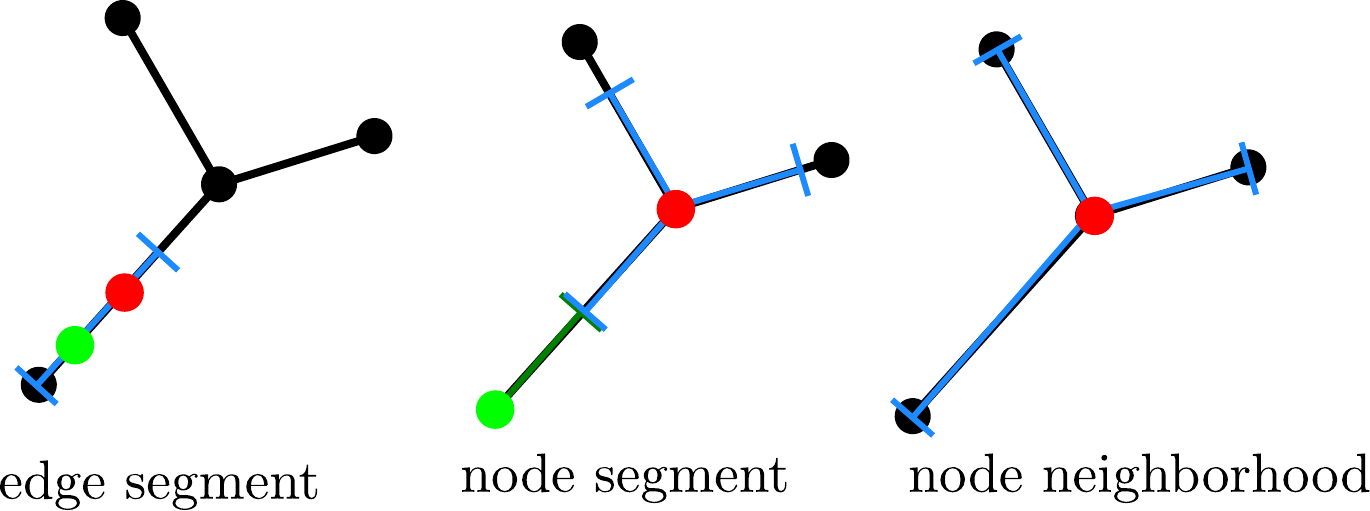}
	\caption{Protective domains used in multi-particle simulations. The network region is shown in black, the particle of interest in red, and the protective domain containing it in blue. An edge segment domain is confined to a single edge and can contain a pair of particles (green). A node segment domain is centered on a node, with equal-length segments along all edges; it is employed when a nearby particle with its own domain (green) prevents a direct hop onto an adjacent node. A node neighborhood domain encompasses the full edges connected to a given node, and requires the particle to start on the node itself.}
	\label{fig:domains}	
\end{figure}

In our system, a protective domain around a particle has one of the following three forms (Fig.~\ref{fig:domains}). An edge segment domain is a single linear segment with end-points on one network edge. This domain can contain one or two particles and is made as big as possible given the edge length and the presence of any additional particles. A node segment domain is centered on a node and surrounded by equal-length segments of the adjacent edges. The segment lengths are set to the minimum of the distance to the nearest unoccupied node or half the distance to the nearest other particle along each edge. Finally, the full node neighborhood domain is employed in the case where there are no other particles on the nodes or edges adjacent to the current node. The asynchronous time propagation for multiple particles proceeds as described in~\cite{oppelstrup2009first} and summarized below.

At the start of the simulation, all particles are distributed into non-overlapping protective domains of the form illustrated in Fig.~\ref{fig:domains}.
The particles all begin synchronously with the global time set to zero. We then proceed to sample the first passage time for each particle or pair of particles to leave its protective domain, as well as the boundary through which they leave. This results in an ordered queue of times $t_i< \ldots < t_k$ for leaving the domain. The particle with the shortest leaving time $t_i$ is propagated to one of the edges of its domain. Any other particle ($j_1, j_2$, etc) whose domain touches the boundary reached by particle $i$ is also propagated forward to time $t_i$, using the spatial propagator appropriate to its domain (eg: Eqs.~\ref{eq:Yikinv},~\ref{eq:Yedge}). The affected particles $i, j_1, j_2, \ldots$ are then partitioned into protective domains. New exit times are selected for these particles and added into the priority queue.
 Each step of the algorithm thus consists of propagating a single particle using the appropriate first-passage time distribution for its domain and one or more particles using the spatial ``no-passage" distribution within their domain. The number of such no-passage particles is at most equal to two (if particle $i$ ends on an edge) or the degree of the network node to which it has transitioned.

The first-passage and no-passage propagators for a single particle have already been derived for a node-neighborhood domain (Eqs.~\ref{eq:Pikinvert} and~\ref{eq:Yikinv}, respectively) and an edge-segment domain (Eq.~\ref{eq:Sedge} and~\ref{eq:Yedge}). A node-segment domain, consisting of $d$ segments of equal length $\ell$, can be treated identically to a single one dimensional interval with absorbing boundaries at $(-\ell,\ell)$ and the particle starting at $x_0 = 0$. Given the symmetry of the domain, the particle has a $1/d$ chance of ending a first-passage or a no-passage step on any one of the edges. The joint propagation of a pair of particles that share an edge segment is described in~\ref{app:pairpropagation}.

\begin{figure}
	 \includegraphics[width=0.5\textwidth]{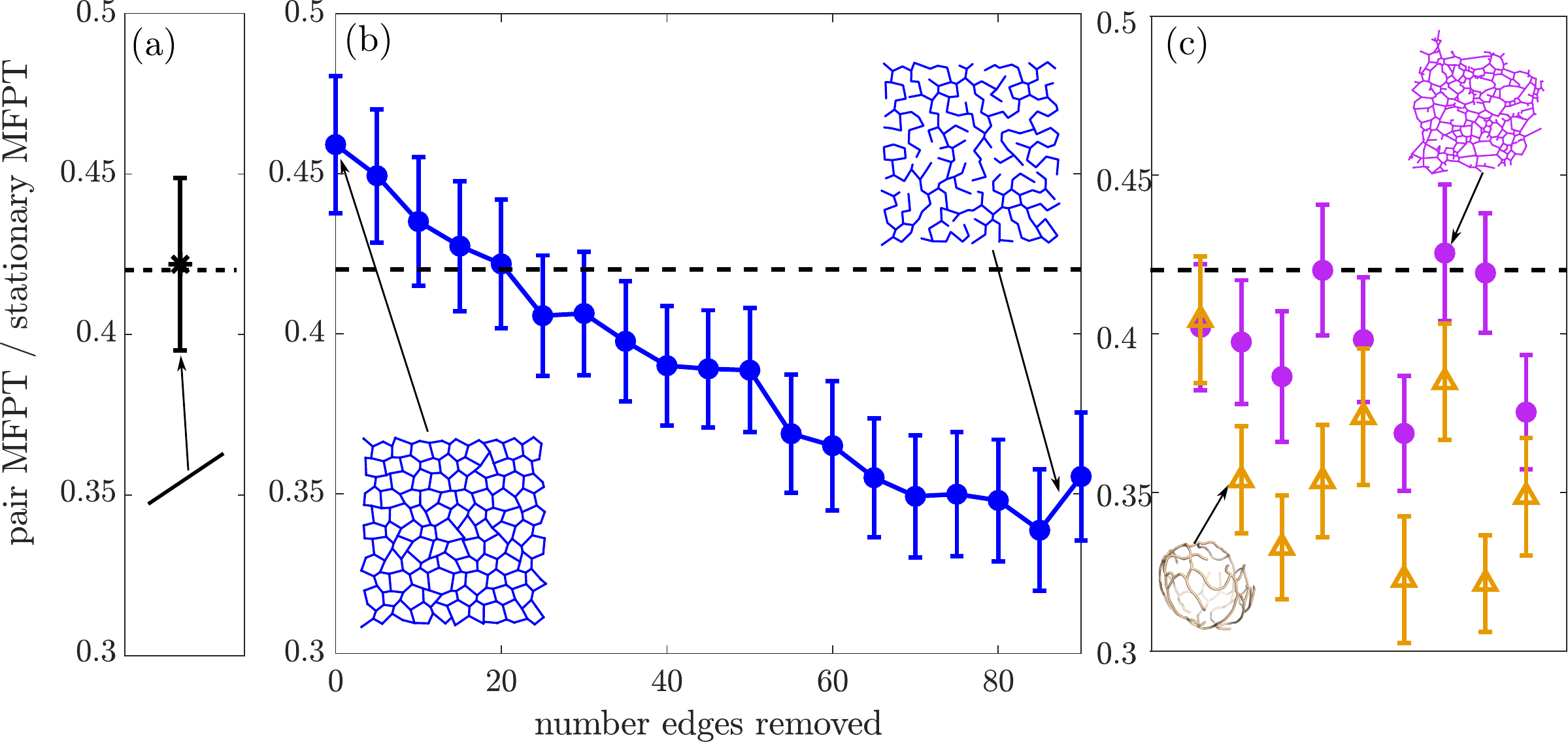}
	\caption{First encounter times for two diffusing particles (with same diffusivity), compared to the analytic tagMFPT to a stationary target. Target initial position is averaged over all network nodes, and particles start uniformly distributed along network edges. (a) Relative search time on network consisting of single line of nodes. Dashed black line indicates analytic expected value on a finite-length linear segment. (b)
		 Relative search times on a lattice-like synthetic network with increasing number of edges removed, reducing connectivity.
		  (c) Relative search times on 9 ER network structures from COS7 cells (magenta circles), 9 yeast mitochondrial network structures (orange triangles). Error bars show standard error of the mean, for simulations of 500 particles. }
	\label{fig:pairsims}	
\end{figure}

\subsubsection{Example: Pair encounter times}
We use simulations on a variety of network structures to compare the encounter time for two diffusive particles ($\tau_\text{enc}$) versus the mean first passage of a single diffusive particle searching for a stationary target placed anywere on the network (target-averaged global MFPT, or tagMFPT). On an infinite line or plane, the separation between a pair of particles is itself diffusive, with an effective diffusivity $2D$. The pair encounter time is thus expected to be half that of the MFPT to a stationary target. In a finite domain, the pair encounter time is dependent on the initial separation between the two particles, as well as on their position relative to the domain boundaries. An exact solution for this encounter time is available for pair diffusion on a line segment~\cite{tejedor2011encounter}, where averaging over starting positions gives
\begin{equation*}
\begin{split}
\tau_\text{enc}^{(1D)} &= \frac{L^2}{D} \( \frac{1}{8} + \frac{32}{\pi^5 } \sum_{k=0}^{\infty} \frac{\cos\(k\pi\)-\sinh\(\(k+\frac{1}{2}\) \pi\) }{\(2k+1\)^5 \cosh\( \(k+\frac{1}{2} \) \pi\) } \) \\%
	 &\approx 0.42 \times \text{ tagMFPT}.
\end{split}
\end{equation*}
Simulation results using our algorithm reproduce this expected behavior on a simple network consisting of a line of nodes (Fig.~\ref{fig:pairsims}a). We next look at pair particle behavior on a planar lattice-like network structure with degree 3 nodes and multiple edges removed to reduce network connectivity. The edges are chosen at random while keeping a single connected cluster within the network. For a full lattice network, the pair encounter time is nearly half of the tagMFPT. However, as the network connectivity decreases and approaches a tree-like loop-less structure, the pair encounter time drops substantially lower, to about $0.3 \text{ tagMFPT}$ (Fig.~\ref{fig:pairsims}b). On poorly connected networks the ability of a target to move diffusively can thus greatly speed up the encounter time, beyond what would be expected in a bulk continuum geometry. This result stems from the fact that poorly connected networks have highly heterogeneous global mean first passage times, depending on where a stationary target is placed~\cite{brown2020impact}. If the target is located in regions of the network that are difficult to access, the MFPT can be several times higher than the average over all target positions~\cite{brown2020impact}. Even when averaging over different target locations, the extremely poor accessibility of some target positions leads to tagMFPTs that are more than twice as slow as the pair encounter time where both particles are diffusive and capable of leaving the hard-to-access regions. 

We also compare the pair encounter time to the stationary target search time for several organelle network structures (Fig.~\ref{fig:pairsims}c). Given their high connectivity, peripheral ER network structures have a pair encounter time that is, on average, $\left<\tau_\text{enc}^{(ER)} / \text{tagMFPT} \right> \approx 0.40\pm 0.007$ (standard error over networks). Even for these highly looped structures, the pair encounter time is less than half the averaged first passage time to a stationary target. Yeast mitochondrial network structures are less well connected (less looped) than the mammalian ER~\cite{brown2020impact}, and their pair encounter time is slightly lower: $\left<\tau_\text{enc}^{\text{(mito)}}/\text{tagMFPT} \right> \approx 0.35\pm 0.009 $.
These results imply that diffusion of ER and mitochondrial structures such as ERES components or nucleoids may be helpful in allowing for rapid search by diffusing proteins.

\section{Discussion}

The methodology outlined in this work allows for precise modeling of the propagation of diffusive particles confined in tubular network structures. We first describe analytic calculations for the mean and variance of the first passage time to reach one of several stationary target sites.  Our approach is distinct from past studies of transport on spatial networks~\cite{masuda2017random, katifori2010damage, grebenkov2018heterogeneous, dora2020active} by explicitly considering the physical process of diffusion between network nodes, allowing the exact calculation of appropriate splitting probabilities and mean hopping times between adjacent nodes, as a function of the relevant edge lengths. Applications to peripheral ER network structures indicate the effective kinetics in the ER are intermediate between those expected in a 1D \emph{vs.}\ 2D geometry (Fig.~\ref{fig:targetdensity}), as well as highlighting the spatial heterogeneity in first passage times between different regions of a network (Fig.~\ref{fig:variance}).

The analytic calculations are then further expanded to consider partially reactive edges scattered throughout the network, giving spatially heterogeneous reaction rates. This model allows an exploration of how the spatial distribution of a fixed total absorbance affects reaction kinetics. In a single long tube, we show that spreading out an absorbing region over an intermediate length-scale results in optimal mean reaction times (Fig.~\ref{fig:1Dabsorb}). This result is applied to distal cargo capture by microtubule tips and implies a benefit to heterogeneous microtubule lengths in extended cell domains such as neuronal axons and fungal hyphae. In a two-dimensional lattice-like network, dispersal of partially absorbing regions away from the source of particle production is shown to be advantageous primarily when particles have a finite maturation time period before they can react (Fig.~\ref{fig:donut}). While protein production in the endoplasmic reticulum is thought to occur largely in the perinuclear rough ER, secreted proteins must undergo a folding and maturation process over timescales of minutes to hours before they can exit the network~\cite{hebert2007inandout}. Our results thus indicate a potential functional role for the dispersion of ER exit sites throughout the periphery, where they can catch diffusing proteins as they mature. 

Complementing the analytic MFPT calculations, we also present an exact, efficient algorithm for agent-based reaction-diffusion simulations of particles on a tubular network. This algorithm constitutes a special case of kinetic Monte Carlo~\cite{oppelstrup2009first} or Green's Function Reaction Dynamics~\cite{van2005simulating} approaches, tailored for a network geometry. Unlike classic Brownian Dynamics, it allows for step sizes on the order of entire edge lengths, while sampling from the exact node propagator distribution function. 
In contrast to a variety of prior studies of stochastic trajectories on networks~\cite{noh2004random, colizza2007reaction,dora2020active}, 
we do not assume a single well-defined timescale for each node transition and instead incorporate the full distribution of transition times and splitting probabilities based on physical diffusion over edges of arbitrary length. Such simulations allow for calculations that go beyond the mean target search time, to explore reaction time distributions (Fig.~\ref{fig:simvalidate}), extreme particle statistics (Fig.~\ref{fig:extremevals}),  spatial dispersion (Fig.~\ref{fig:MSDsims}), and pair encounter times between multiple diffusive particles (Fig.~\ref{fig:pairsims}). Interestingly, the pair encounter times are found to be less than half the timescale for hitting a stationary target, particularly in the case of poorly-connected networks. This finding implies that allowing targets to become diffusively mobile, rather than fixed to a specific network region, can significantly increase their encounter rates with other diffusing particles. 

 Our approach is relevant to any spatial network that can be described by curved one-dimensional edges (tubules) connecting nodes of negligible volume. For realistic reticulated networks the implied assumption is that particles spend a negligible amount of time in tubule junctions and are equally likely to enter any of the edges emerging from the junction. While mitochondrial networks generally have nodes that appear to simply be junctions of joining edges~\cite{viana2020mitochondrial}, the endoplasmic reticulum may include more complex node morphologies such as small regions of fenestrated sheets~\cite{schroeder2019dynamic}.
While our methods do not currently describe diffusion within such peripheral sheets, they can incorporate the dense patches of tubular intersections that have been hypothesized to comprise some such regions~\cite{nixon2016increased}.  
  Including non-negligible trapping times within expanded nodal regions could serve as an important avenue for future work in modeling diffusive transport throughout the ER.

Another potential source of complication in cellular organelles is the combination of diffusive transport with other dynamic processes such as network structural remodeling 
and the recently postulated intra-tubule flows that may contribute to rapid motion of ER luminal proteins~\cite{holcman2018single, dora2020active}. Topological rearrangements of both mitochondria and the ER occur on the tens of seconds to minutes timescales~\cite{rafelski2013mitochondrial,friedman2011er,lin2014structure}. Network remodeling is thus unlikely to have a substantial effect on local protein propagation but may modulate the long search times to reach poorly connected regions of the network. Putative flows in the ER lumen have been proposed to rapidly drive luminal proteins betweeen adjoining nodes, without affecting the diffusion of membrane-embedded proteins~\cite{holcman2018single}. Simulations incorporating such flows would require modifying the local node-to-node propagator function
for diffusion with drift along individual edges. Both time-varying flows and network rearrangements would require propagating particles forward to synchronization times when such network-level changes occur. In the case of localized changes (ie: flow reversal along a given edge, or extension of a single tubule), substantial speed-up could still be obtained using the asynchronous kinetic Monte Carlo approach described above.

The methodology presented here has broad applicability to modeling the diffusion of particles over networks with physical edges. Our examples focus on applications to  intracellular organelle networks, forming a mathematical framework for exploring diffusion-limited reaction processes in reticulated mitochondria and the ER. These two organelles host a variety of biomolecular pathways of interest, and understanding the kinetics of diffusive search is critical to investigating dynamics of such disparate cellular processes as the early secretory pathway (ER), calcium release and replenishment (ER), and mitochondrial gene transcription.
 The described framework for simulation and analytic computation of diffusive reaction times in reticulated structures forms an important foundation for exploring the link between  morphology and function in complex biological architectures.

\begin{acknowledgements}
	The authors thank Dr. Matheus Viana for providing access to the yeast mitochondrial network structure database. Funding for this work was provided by the National Science Foundation (Grant ID \#2034482, to EFK and LMW), a Cottrell Scholar Award from the Research Corporation for Science Advancement (EFK), and the Hellman Fellows Fund (EFK). 
\end{acknowledgements}

\bibliographystyle{unsrtnat}
\bibliography{propagatormethods}

\appendix

\section{Code availability}
Example Matlab code for carrying out the calculations described in this manuscript (for both analytic calculations and simulations) is available at \url{https://github.com/lenafabr/propagatormethods}. The details of the network construction and simulation algorithms are outlined in the appendices below.

\section{Generating network structures}
\label{app:networks}

\subsection{Cellular network structures}
Peripheral ER network structures were obtained from confocal imaging of live COS7 (monkey kidney) cells, labeled with KDEL venus as a fluorescent luminal marker. Cell culture and imaging was carried out exactly as described in prior work~\cite{brown2020impact}. Images of nine individual cells were analyzed using Ilastik, an interactive machine-learning tool~\cite{berg2019ilastik}, to classify pixels as belonging to the ER network structure. The images were then segmented (with standard Otsu thresholding) and skeletonized using built-in algorithms in Matlab~\cite{MATLAB:2018}. The network topology was extracted by grouping skeleton nodes in adjacent pictures and using Matlab's bwtraceboundary subroutine to trace out skeleton boundaries between groups of nodes.

Mitochondrial networks in yeast cells were taken directly from published data~\cite{viana2020mitochondrial} available at \url{https://data.mendeley.com/datasets/nshn8hhd6d/1}.

\subsection{Synthetic network structures}

To create the synthetic peripheral ER networks featured in Fig.~\ref{fig:donut}, we began with a regular honeycomb lattice, $30$ lattice cells across, and excised an annular region. The inner and outer radii of the anulus were chosen to correspond with the dimensions of the nucleus ($8\mu$m) and the cell boundary ($20\mu$m) for a typical COS7 cell. The network edge length is $0.8\mu$m, matching the average edge length ($0.78\pm0.02\mu$m, mean plus or minus standard error of the mean) computed for the extracted ER networks used in this study.

The synthetic networks from Figs.~\ref{fig:simvalidate},~\ref{fig:MSDsims}, and~\ref{fig:pairsims} were also initiated as honeycomb lattice networks, 10 lattice cells across. The spatial positions of the nodes were then randomly perturbed by an amount $\delta d \in(0,0.2\ell)$, with $\ell$ the minimum edge length on the network. For fig.~\ref{fig:pairsims}, random sets of edges were selected for removal, such that the network retained a single connected component with each subsequent edge removed. The results are averaged over 10 different network structures generated by removing distinct sets of edges.

All edges in the synthetic networks were treated as straight lines between connected nodes.

\section{Sampling transition times in simulations}
\label{app:sampling}

\subsection{Transition times between node neighborhoods}

For a particle that starts at node $i$, we sample the distribution of times $t$ for leaving the neighborhood of $i$ according to the following algorithm. 

First, we select the next node reached by the particle, using the discrete splitting probabilities $P_{ik}^*$ (Eq.~\ref{eq:P*}). Next, consider the distribution of survival times $S_{ik}(t)$, conditional on the particle eventually leaving the neighborhood through node $k$:
\begin{equation}
\begin{split}
S_{ik}(t) & = \frac{1}{P_{ik}^*}\int_t^\infty P_{ik}(t') dt'  = \frac{1}{P_{ik}^*}\sum_p \frac{r_p^{(ik)}}{Du_p^2} e^{-Du_p^2 t}.
\end{split}
\label{eq:cumleave}
\end{equation}
The roots $u_p$ for the Laplace inversion are precalculated for each neighborhood prior to the start of the simulation. The infinite summation is truncated to $u_p < u_\text{max}$, with the maximum root for each neighborhood selected as described in~\ref{sec:shortt}.

Sampling of the transition time is done through the inverse cumulant method, by first selecting a uniformly distributed value $w \in (0,1)$ and then numerically solving for the value of $t$ with $S_{ik}(t)= w$. 

In practice, we pre-tabulate the distributions of leaving times $S_{ik}(t)$ for each node and each adjacent edge in a network, allowing for rapid sampling of many particle transitions initiating at that node.

\subsection{Short time asymptotic limit}
\label{sec:shortt}
Eq.~\ref{eq:Pikinvert} for the Laplace inversion of the leaving time distribution converges quickly for long times and short edges, with the number of required summation terms scaling as $\ell^2/(Dt)$. In the limit of short times, an alternate approach is taken for the Laplace inversion. Specifically, we expand Eq.~\ref{eq:flux} in the limit of $s\rightarrow \infty$ to get the lowest order terms
\begin{equation}
\begin{split}
\widehat{P}_{ik}  \xrightarrow{\alpha \rightarrow \infty} & \frac{2}{n} \(e^{-\alpha \ell_{ik}} + e^{-3\alpha \ell_{ik}} \) \\
& -\frac{4}{n^2}
   \sum_{j=1}^n e^{-\alpha (\ell_{ik} + 2\ell_{ij})}
\end{split}
\label{eq:Psinf}
\end{equation}
where $\alpha = \sqrt{s/D}$ and $n$ is the degree of node $i$.

We then use the relation
\begin{equation}
\mathcal{L}^{-1}\(\frac{1}{s} e^{-2a\sqrt{s}}\) = \text{Erfc}\(\frac{a}{\sqrt{t}}\)
\end{equation}
to invert the individual terms of Eq.~\ref{eq:Psinf}, integrated over time. 
This yields the final expression for the cumulative conditional probability that a particle leaves the neighborhood of node $i$ by time $t$, given it eventually leaves to node $k$:
\begin{equation}
\begin{split}
1 - S_{ik}(t) \xrightarrow{t \rightarrow 0} & \frac{2}{nP_{ik}^*}\(\text{Erfc}\(\frac{\ell_{ik}}{2\sqrt{Dt}}\) + \text{Erfc}\(\frac{3\ell_{ik}}{2\sqrt{Dt}}\) \) \\ %
&-\frac{4}{n^2P_{ik}^*} \sum_j \text{Erfc} \(\frac{\ell_{ik}+2\ell_{ij}}{2\sqrt{Dt}}\)
\end{split}
\label{eq:shortt}
\end{equation}

In order for this limit to be accurate, the next highest term in the expansion (Eq.~\ref{eq:Psinf}) needs to be sufficiently small. The next highest term for any given time is determined by the shortest edge length ($\ell_{i,\text{min}}$) and must be at most $\text{Erfc}\left[5\ell_{i,\text{min}}/(2\sqrt{Dt})\right]$. Therefore, if we set a tolerance of $\epsilon$,  and define the cutoff $\xi=-\log\epsilon$, then the expression in Eq.~\ref{eq:shortt} is accurate for all times $t$ below
\begin{equation}
\begin{split}
t^* = \frac{25\ell_{i,\text{min}}^2}{4\xi D}.
\end{split}
\end{equation}

For times above $t^*$, we can employ the converging series of Eq.~\ref{eq:Pikinvert}. The series is truncated by considering only those poles that satisfy $Dt^* u_\text{max}^2 > \xi$, or equivalently
\begin{equation*}
\begin{split}
u_\text{max} = \frac{2\xi}{5\ell_\text{min}}.
\end{split}
\end{equation*}

Eq.~\ref{eq:roots} has poles at $m\pi/\ell_{ij}$ for all integer $m$ and a root between every pair of consecutive poles. The maximum number of roots up to $u_\text{max}$ is then given by 
\begin{equation}
\begin{split}
\sum_j m_{ij,\text{max}} = \frac{2 \xi}{5\pi \ell_\text{min}} \sum_j \ell_{ij}.
\end{split}
\end{equation}

The cumulative distribution function can therefore be evaluated efficiently for all values of time, so long as the ratio of longest to shortest edge length in the neighborhood is not too large.

An analogous approach can be used to find the no-passage spatial propagation of a particle over a very short time interval, conditional on never leaving the neighborhood of the node. For a particle starting on node $i$ that has never passed any of the adjacent nodes, the probability that after time $t$ it is located on edge $ik$, between $x$ and $\ell_{ik}$ can be derived from Eq.~\ref{eq:Yik} as
\begin{equation}
\begin{split}
\widehat{Y}_{ik}(x,s) &  \xrightarrow{\alpha \rightarrow \infty} \frac{2}{sn^2} 
\left[1 - e^{-\alpha(\ell_{ik}-x)}\right] \sum_{j=1}^n e^{-2\alpha \ell_{ik}} \\
Y_{ik}(x,t) &  \xrightarrow{t \rightarrow 0} \frac{2}{n^2} \sum_j 
\left[\text{Erfc}\(\frac{\ell_{ij}}{\sqrt{Dt}}\) \right.\\
 & \quad \quad - \left.\text{Erfc}\(\frac{2\ell_{ij}+\ell_{ik}-x}{\sqrt{Dt}}\)\right]
\end{split}
\label{eq:Yshortt}
\end{equation}
This expression is normalized by $Y_{ik}(0,\Delta t)$, and allows for a sampling of the $x$ coordinate using the inverse cumulant method.

\subsection{Propagation on edges}

For a particle starting at position $x_0$ on edge $ik$, the survival probability to time $t$ given that it eventually leaves at one of the bounding nodes can be obtained directly via Eqs.~\ref{eq:cumleave},~\ref{eq:shortt} by treating the starting position as its own node with adjacent nodes $i$ and $k$ connected by edges of length $\ell_{ik}-x_0$ and $x_0$, respectively. Analogously, Eqs.~\ref{eq:Yikinv},~\ref{eq:Yshortt} can be leveraged to sample the spatial propagation along an edge over a time interval $\delta t$ until the next save-point. 

For this special case, the splitting probability and survival probability over time match to the known solutions for a particle on a one-dimensional interval with two absorbing boundaries~\cite{redner2001guide,oppelstrup2009first}. Namely, for a particle beginning at position $x_0$ on an edge of length $\ell_m$, we have
\begin{equation}
\begin{split}
P_i & = x_0/\ell_m, \quad P_k = 1 - x_0/\ell_m, \\
S^{(E)}_{m,+}(t) & = \frac{2 \ell_m }{\pi x_0} \sum_{p=1}^\infty  \frac{(-1)^{(p+1)}}{p}\sin\(\frac{p\pi x_0}{L}\) e^{-\frac{p^2\pi^2 D t}{\ell_m^2}}, \\
S^{(E)}_{m,-}(t) & = \frac{2 \ell_m }{\pi(\ell_m-x_0)} \sum_{p=1}^\infty  \frac{1}{p}\sin\(\frac{p\pi x_0}{L}\) e^{-\frac{p^2\pi^2 D t}{\ell_m^2}},
\end{split}
\label{eq:Sedge}
\end{equation}
where $S^{(E)}_{m,\pm}(t)$ correspond to the probability the particle has not left by time $t$, given it eventually leaves through the boundaries at $\ell_m$ and $0$, respectively. Similarly, the no-passage spatial propagator for the particle on an edge can be expressed as,
\begin{equation}
\begin{split}
Y^{(E)}_{m}(x, t) & = \frac{2}{\pi} \sum_{p=1}^\infty  \frac{1}{p}\sin\(\frac{p\pi x_0}{\ell_m}\) \\
& \times \left[1 - \cos\(\frac{p\pi x}{\ell_m}\)\right] e^{-\frac{p^2\pi^2 D  t}{\ell_m^2}}.
\end{split}
\label{eq:Yedge}
\end{equation}

For the short-time limit, analogous functions for a particle propagating on an edge can be derived from Eqs.~\ref{eq:shortt} and~\ref{eq:Yshortt}.

\subsection{Particle pair propagation}
\label{app:pairpropagation}
In order to avoid infinitesimally small steps when particles approach each other, it is necessary to enable two particles to share a single domain, where particle encounter is a finite-probability event within the domain. We follow the algorithm for paired particle propagation that was outlined previously in~\cite{oppelstrup2009first}. The essential points of the algorithm are presented here for completeness, and generalized to particles with arbitrary diffusivity.

Specifically, assume $x<y$ describe the coordinates of two particles along a segment with absorbing boundaries at $0$ and $L$. If we make the coordinate change defined by
\begin{equation}
\begin{split}
u & = x -y \\
v & = \sqrt{\frac{D_y}{D_x}} x + \sqrt{\frac{D_x}{D_y}} y,
\end{split}
\end{equation}
then the time evolution for the joint distribution of the two particles can be expressed as a two-dimensional diffusion equation~\cite{ambjornsson2008single},
\begin{equation}
\begin{split}
\frac{\partial c(u,v)}{\partial t} = (D_x+D_y) \left[\frac{\partial^2 c}{\partial u^2} +  \frac{\partial^2 c}{\partial v^2}\right].
\end{split}
\end{equation}
The joint evolution of the transformed variables can thus be treated as two independent 1D diffusion processes, both with diffusivity $D_x+D_y$.

\begin{figure}
	\includegraphics[width=0.4\textwidth]{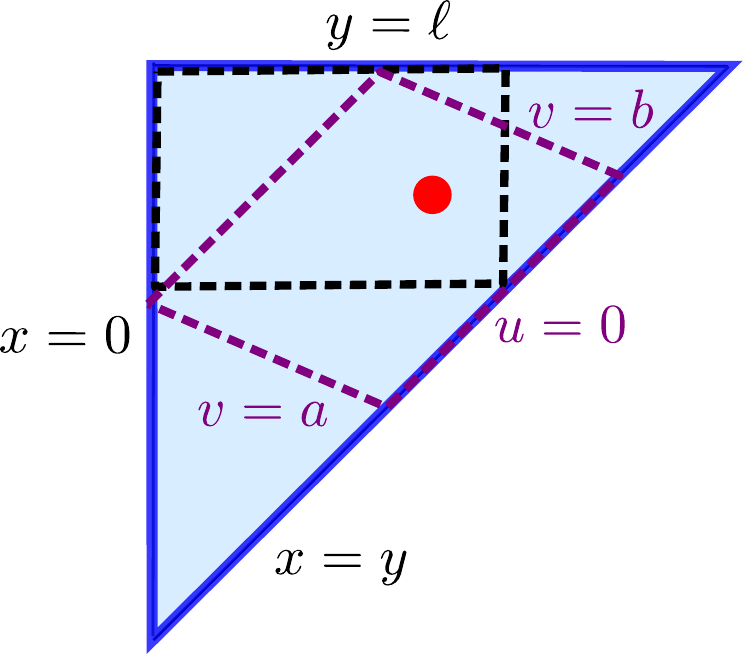}
	\caption{Propagators used for an edge-segment protective domain containing a pair of particles. Blue lines mark absorbing boundaries in the domain. Red dot marks $x_0, y_0$, the initial positions of the two particles. Two propagation approaches are possible. The $x$ and $y$ coordinates can be propagated directly in an inscribed rectangular domain (black dashed lines). Alternately, the transformed coordinates $u, v$ can be propagated in the region outlined in purple, allowing for particle encounter ($u=0$) as a finite-probability outcome of the propagation step. The choice of propagation is based on greatest minimum distance to the boundary -- in this case, the purple region would be used. }
	\label{fig:pairpropagator}	
\end{figure}

The edge segment domain containing a pair of particles is illustrated in Fig.~\ref{fig:pairpropagator}, which highlights the three boundaries for exiting the domain: $x=0$ (first particle hits a segment boundary), $y=L$ (second particle hits a segment boundary), or $x=y$ (particles encounter each other). For simplicity of sampling we propagate the system forward within a smaller region inscribed within this two-dimensional domain. One option is to draw a rectangle (black dashed line in the figure), aligned with the $x$ and $y$ axes contained within the allowed domain. The particles are then propagated individually to the boundaries of the rectangle, using Eq.~\ref{eq:Sedge} to find the earliest time for either particle to hit a boundary and Eq.~\ref{eq:Yedge} to spatially propagate the other particle on the condition that it does not leave the rectangle. This approach, however, does not allow the particles to encounter each other and results in infinitesimally small time steps as the particles come close together. An alternate approach is to inscribe a parallelogram within the domain that has one edge on the $x=y$ line (corresponding to $u=0$) and the other edges parallel to the $v$ axis, along $v=a$ and $v=b$. The coordinates $u$ and $v$ are propagated diffusively within this parallelogram according to Eq.~\ref{eq:Sedge} and~\ref{eq:Yedge}, and the two particle positions $x, y$ are then recalculated accordingly. If the exit from the parallelogram happens along the line $u=0$ then the particles encounter each other before leaving the domain and a reaction event is recorded. 

The choice between the two subdomains (rectangle in $x,y$ coordinates or in $u,v$ coordinates) is made by picking the one with the biggest minimum distance of the initial pair position to the boundaries of the domain. In other words, when the particles start close together ($u_0$ close to $0$), the $u, v$ coordinate system is selected to allow an encounter to occur. When they are further apart, the $x, y$ coordinate system is used to allow larger time steps in propagating to a domain boundary.

It should be noted that the time steps in this algorithm can get very small when two particles approach each other in the vicinity of a node, resulting in $x$ and $y$ both close to $0$. However, our propagation approach allows for a substantial probability of either the particles encountering each other or one of the particles moving away from the node on a different segment, so that this unfavorable case does not persist for many time steps and does not substantially slow down the simulations when particle density is sparse.

\end{document}